\documentclass[aps,pra,amsmath,amssymb,floatfix,twocolumn,amsmath,superscriptaddress,twocolumn,nofootinbib,tighten,letterpaper]{revtex4}
\usepackage{multirow}
\usepackage{bbold}
\usepackage{subfigure}
\usepackage{color}
\usepackage{mathrsfs}
\usepackage{hyperref}
\usepackage[normalem]{ulem}
\usepackage{bm}

\usepackage{amsfonts, relsize, color}
\usepackage{graphics}
\usepackage{graphicx}
\usepackage{subfigure}
\usepackage{hyperref}
\usepackage{color}

\begin{document}
\title{\bf Itinerant quantum multicriticality of two-dimensional Dirac fermions}

\author{Bitan Roy}
\affiliation{Max-Planck-Institut f$\ddot{\mbox{u}}$r Physik komplexer Systeme, N$\ddot{\mbox{o}}$thnitzer Str. 38, 01187 Dresden, Germany}

\author{Pallab Goswami}
\affiliation{Department of Physics and Astronomy, Northwestern University, 2145 Sheridan Road, Evanston, IL 60208, USA}

\author{Vladimir Juri\v ci\' c}
\affiliation{Nordita,   KTH Royal Institute of Technology and Stockholm University, Roslagstullsbacken 23,  10691 Stockholm,  Sweden}

\date{\today}
\begin{abstract}
We analyze emergent quantum multicriticality for strongly interacting, massless Dirac fermions in two spatial dimensions ($d=2$) within the framework of Gross-Neveu-Yukawa models, by considering the competing order parameters that give rise to fully gapped (insulating or superconducting) ground states. We focus only on those competing orders, which can be rotated into each other by generators of an exact or emergent chiral symmetry of massless Dirac fermions, and break $O(S_1)$ and $O(S_2)$ symmetries in the ordered phase. Performing a renormalization group analysis by using the $\epsilon=(3-d)$ expansion scheme, we show that all the coupling constants in the critical hyperplane flow toward a new attractive fixed point, supporting an \emph{enlarged} $O(S_1+S_2)$ chiral symmetry. Such a fixed point acts as an exotic quantum multicritical point (MCP), governing the \emph{continuous} semimetal-insulator as well as insulator-insulator (for example, antiferromagnet to valence bond solid) quantum phase transitions. In comparison with the lower symmetric semimetal-insulator quantum critical points, possessing either $O(S_1)$ or $O(S_2)$ chiral symmetry, the MCP displays enhanced correlation length exponents, and anomalous scaling dimensions for both fermionic and bosonic fields. We discuss the scaling properties of the ratio of bosonic and fermionic masses, and the increased dc resistivity at the MCP. By computing the scaling dimensions of different local fermion bilinears in the particle-hole channel, we establish that most of the four fermion operators or generalized density-density correlation functions display faster power-law decays at the MCP compared to the free fermion and lower symmetric itinerant quantum critical points. Possible generalization of this scenario to higher-dimensional Dirac fermions is also outlined.
\end{abstract}

\maketitle

\section{Introduction}

Generally a global symmetry (present at the microscopic level) broken inside an ordered state gets restored at the critical point separating the ordered and the disordered phases. Such a critical point supports fluctuations of the order parameter at all length scales, giving rise to fascinating scaling properties for various physical quantities. More intriguingly, a class of critical points can also possess different emergent symmetries, which are otherwise absent at the microscopic level. For example, consider the coupling between massless Dirac fermions and gapless, relativistic bosons with different velocities at the microscopic level or high-energy scales, where the bosonic field can describe either a fluctuating order parameter or a $U(1)$ gauge field. At the renormalization-group (RG) fixed point for such a theory, the velocities for both fermionic and bosonic degrees of freedom become equal, and the difference between two velocities can act as an irrelevant perturbation, leading to the notion of an emergent \emph{Lorentz invariance} in the system~\cite{nielsen-chadha-lorentz, vozmediano-lorentz,  sslee-lorentz, anber-lorentz, nagaosa-loretz, Bednik-lorentz, roy-lorentz, goswami-raghu}.

However, the notion of emergent symmetry for RG fixed points is not limited to the Lorentz invariance. A fascinating yet simple example arises in the context of topological quantum phase transitions for noninteracting fermions or Hartree-Fock quasiparticles of an ordered state. For a large number of symmetry classes, the quantum phase transition between topological and trivial insulating or superconducting phases occurs through the closing of a spectral gap, and the quantum critical properties are described by massless Dirac fermion fixed points at different spatial dimensions. At the microscopic level, when the spectral gap is closed, the underlying Dirac fermion still possesses a momentum dependent Wilson mass (quadratic or higher order in momentum), carrying the information regarding the underlying symmetry class. But, at the RG fixed point governing the low energy properties, such momentum dependent mass becomes an \emph{irrelevant perturbation} in comparison to the $k$-linear dispersion (characteristic feature of Dirac fermions). These massless Dirac fermion fixed points exhibit at least an emergent $U(1)$ chiral symmetry, which is absent in the microscopic theory of  symmetry classes~\cite{goswami-chakravarty-1, roy-goswami-sau, goswami-chakravarty-2, roy-alavirad-sau}. Intriguingly, such a massless Dirac fixed point in one dimension also controls the critical properties of a transverse field Ising chain~\cite{mattis, kogut-rmp, sachdev}.

More exotic examples of emergent chiral symmetry arise for the gapless phase and critical points for spin-1/2 chains in one dimension~\cite{sachdev,tsvelik}. For an one dimensional spin-1/2 chain, the components of O(3) antiferromagnetic order parameter and the Z$_2$ valence bond solid order parameter can be rotated into each other by a chiral transformation. In the SU(2) symmetric algebraic spin liquid phase of isotropic, nearest neighbor antiferromagnetic Heisenberg model for the spin-1/2 chain, none of these order parameters can condense, and an emergent O(4) chiral symmetry for massless Dirac spinons is realized~\cite{haldaneaffleck}. However, for suitable anisotropic exchange couplings in a J$_1$-J$_2$ model, one can find a line of unconventional quantum critical points, which separates an Ising antiferromagnet from a valence bond solid state, which break distinct symmetries~\cite{haldane82}. At the level of effective field theory, the quantum phase transition between these states is accessed by tuning the sign of sufficiently strong umklapp coupling for linearly dispersing spinon excitations. When the umklapp coupling vanishes at the line of critical points, one finds an enhancement of chiral symmetry from Z$_2$ to U(1)~\cite{haldane82,sachdev,tsvelik}. We also note that in the phase diagram of antiferromagnetic J$_1$-J$_2$ model of spin 1/2 chain with XXZ anisotropy, the gapless spin liquid, the Ising antiferromagnet, and the valence bond solid phases meet at a multicritical point (MCP), possessing both emergent Lorentz invariance and $O(4)$ chiral symmetry.

\begin{figure}[t!]
\includegraphics[width=8cm,height=7cm]{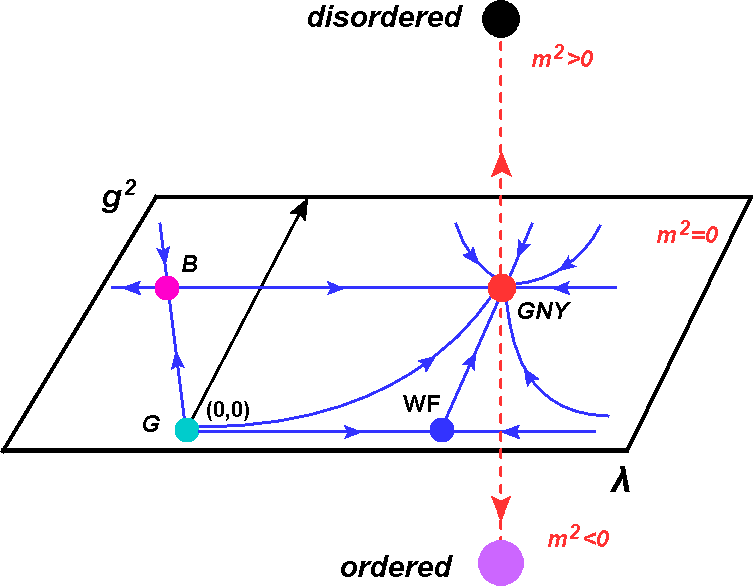}
\caption{ A schematic renormalization group flow diagram of Gross-Neveu-Yukawa (GNY) model. All coupling constants ($g^2$ and $\lambda$) are measured in units of $\epsilon$, the deviation from the upper-critical three spatial dimensions with $\epsilon=3-d$. Notice that in the critical $m^2=0$ plane, the Gaussian (G) fixed point (shown in dark cyan) is fully unstable, and the Wilson-Fisher (WF) fixed point (shown in blue), controlling a continuous quantum phase transition in a pure bosonic system also becomes unstable due to boson-fermion (Yukawa) coupling ($g^2$). The bi-critical (B) fixed point (shown in magenta) lies in the unphysical regime ($\lambda<0$). A new quantum critical point (QCP), ``GNY QCP" (shown in red) now controls the continuous quantum phase transition from a disordered or semimetallic to an ordered phase. The bosonic mass parameter $m^2$ plays the role of a tuning parameter at $T=0$, and respectively for $m^2>0$ and $m^2<0$ system flows toward disordered or semimetallic (black circle) and strongly coupled ordered (dark magenta circle) phases.
}~\label{GNY_Final}
\end{figure}

An interesting question is whether such enlargement of chiral symmetry can also occur for strongly interacting systems in higher spatial dimensions. It has been argued that a similar enlargement of symmetry is responsible for the continuous quantum phase transition between the valence bond solid and the antiferromagnetic phases of two dimensional, frustrated spin models in insulating systems~\cite{sachdev,Senthiletal2,Senthiletal3,Sandvik,Kaul,Chalker1,Chalker2}. These arguments have been strengthened by considering the ordered, insulating states described in terms of two dimensional, eight component, massive Dirac fermions. Within this approach, one couples the Dirac fermions with the three components of an $O(3)$ nonlinear sigma model field (for antiferromagnetic order parameter) and the two components of Z$_3$ (on a honeycomb lattice) (or Z$_4$ on a square lattice) valence bond solid order parameter field, and the corresponding five fermion bilinears or mass terms mutually anticommute. Therefore, two physically distinct order parameters can be rotated into each other by using a general chiral transformation as in one dimension. At the transition between two ordered states, the $O(5)$ chiral symmetry is realized, and the low energy theory is described by an $O(5)$ nonlinear sigma model~\cite{FisherSenthil,TanakaHu}. After integrating out the massive, Dirac fermions, one also obtains a Berry phase for the $O(5)$ nonlinear sigma model, captured by the level $k$ topological Wess-Zumino-Witten term~\cite{Abanov,FisherSenthil,TanakaHu,FuSachdev}, which is believed to cause deconfined quantum criticality, supporting fractionalized excitations. The notion of chiral rotation has also been applied to Kondo-Heisenberg models for describing the competition among antiferromagnetism, valence bond solid and Kondo singlet formation~\cite{pallab-si,pallab-si-1,pallab-si-2,pallab-si-3}. However, any direct field theoretic calculation of critical properties for these unconventional quantum phase transitions is prohibitively difficult. In addition, if we begin with an interacting model of massless Dirac fermions, the relationship between the itinerant antiferromagnetic and valence bond solid quantum critical points and the possible direct transition between the two ordered insulating states is still unknown.

Motivated by this issue, we investigate a more general question for strongly interacting itinerant fermions. Can there be an interacting quantum MCP where the disordered, parent metallic phase and different ordered states meet (i.e., the metallic state has comparable propensities toward the formation of multiple ordered phases)? Such a multi-critical point would control the transition between two ordered states, as well as the transition from a disordered metal to the ordered phases. We here address this question by focusing on various competing orders, which generate mass gap for Dirac fermions in two spatial dimensions. Single-layer graphene stands as a prototypical system for studying such phenomena using both field theoretic methods and quantum Monte Carlo simulations~\cite{sorella-0, sorella-1, herbut-assaad-1, chandrasekharan, troyer-honeycomb,herbut-assaad-2, sorella-2, hong-yao-NN-honeycomb, kaul-itinerant, sato-hohendler-assaad}. It is important to stress that massless Dirac fermions in a two-dimensional half-filled honeycomb lattice can be unstable toward multiple broken symmetry phases (depending on the relative strength of various finite-range components of the Coulomb interaction) via continuous quantum phase transitions~\cite{herbut, raghu-honerkamp, honerpkapm, herbut-juricic-roy, Roy-herbut-kekule, vozmediano, gonzalez, dagofer-hohendler, review-1}. In addition, often the order-parameters describing distinct broken symmetry phases can be rotated (by generators of emergent or exact chrial symmetry of massless Dirac fermions) into each other~\cite{ruy-hou-mudry-chamon, herbut-isospin, herbut-lu-roy}. These two features turn this system into a fertile ground for investigating (using both field theoretic and numerical techniques) the intriguing confluence of several competing orders and emergent quantum multi-critical phenomena of itinerant fermions. We therefore focus on the setup provided by this system. Our analysis can also be generalized to investigate similar question for other two-dimensional Dirac system, such as $\pi$-flux square lattice~\cite{affleck}.

We here address the question of possible emergent quantum MCPs for interacting Dirac fermions within the framework of the appropriate Gross-Neveu-Yukawa model, which is believed to capture the critical properties of the quantum phase transition from a Dirac semimetal to a broken symmetry phase (insulators or superconductors)~\cite{zinn-justin-book-1, rosenstein, herbut-juricic-vafek, herbut-juricic-roy-SC, roy-yang, sslee, hong-yao-1, machiejko-zarf, knorr, hong-yao-2, roy-mcp, roy-juricic-mcp, classen-MCP, roy-kennett-yang}. This theory treats both order parameter fluctuations (captured by appropriate relativistic bosonic $\Phi^4$ theory) and the Yukawa coupling between gapless fermions and bosons on an equal footing. The critical properties of an interacting, itinerant fixed point in $d=2$ can then be accessed via a controlled $\epsilon$-expansion about $d=3$ with $\epsilon=3-d$. At the itinerant quantum critical point, the bosonic and fermionic degrees of freedom remain strongly coupled and the universality class of the Dirac semimetal-ordered phase transition becomes distinct from the well studied Wilson-Fisher fixed point of purely bosonic theory. A schematic RG flow diagram of the GNY field theory is displayed in Fig.~\ref{GNY_Final}. Within this framework a potential emergent quantum multi-critical behavior can also be addressed by coupling two competing order-parameters (that can be rotated into each other by the chiral transformation) with gapless Dirac fermions, besides their mutual coupling at the quartic order.

Let us first provide a brief summary of our current understanding of this issue. In the past~\cite{roy-mcp}, it has been shown if the electronic interaction among massless Dirac fermions is conducive toward forming two competing ordered phases, breaking discrete $Z_2$ and continuous $O(2)$ chiral symmetries, then in the critical hyperplane (realized by setting all bosonic masses to zero) there exists a quantum MCP possessing an enlarged emergent $O(3)$ chiral symmetry that controls the transition into either of these two phases, as well as the one that spontaneously breaks an $O(3)$ symmetry. Even though such an $O(3)$ symmetric MCP can be found even in the absence of gapless fermions (at least from leading order $\epsilon$-expansion), they substantially improve its stability in a multi-dimensional coupling constant space. In other words, two simple (and decoupled) quantum critical points (QCPs) describing quantum phase transitions from a Dirac semimetal to an ordered phase that spontaneously breaks either $Z_2$ or $O(2)$ symmetry are generically unstable in the direction of a new fixed point with enlarged chiral symmetry. A $Z_2 \otimes O(3)$ symmetric GNY theory has been analyzed in Ref.~\cite{roy-juricic-mcp}, with the exponents (specifically the bosonic and fermionic anomalous dimensions) being identical to the ones found at an $O(4)$ symmetric MCP.
However, a comprehensive understanding of emergent symmetry among competing orders  in an interacting quasi-relativistic system is still lacking, which motivates our current investigation.

We now offer a brief synopsis of our main findings. In the present work, we make an attempt to unify the notion of emergent symmetry among various competing orders and show that for comparable propensities toward the nucleation of $O(S_1)$ and $O(S_2)$ symmetry breaking orderings, there always exists a quantum MCP possessing an enlarged $O(S_1+S_2)$ symmetry.~\footnote{In this notation $Z_2$ symmetry is denoted by $O(1)$.} Irrespective of specific values of $S_1$ and $S_2$, we show that the $O(S_1+S_2)$ symmetric quantum MCP governs the following continuous quantum phase transitions: (1) Dirac semimetal to an $O(S_1)$ symmetry breaking ordered phase, (2) Dirac semimetal to an $O(S_2)$ symmetry breaking ordered phase, (3) Dirac semimetal to an $O(S_1+S_2)$ symmetry breaking ordered phase, and most strikingly (4) the direct transition between $O(S_1)$ and $O(S_2)$ symmetry breaking ordered phases. All these transitions are characterized by same correlation length exponent [see $\nu^{(S)}$ in Eq.~(\ref{CLE:final})], manifesting emergent $O(S_1+S_2)$ symmetry, and the dynamic scaling exponent $z=1$ (reflecting the restoration of Lorentz symmetry).

Since we restrict our analysis to graphene-like systems, in which low-energy fermionic qusiparticles are described by an eight-component massless Dirac fermion (due to two sublattices, two valleys and two projections of spin) the physically relevant cases are the following:

1. A $Z_2 \otimes Z_2$ symmetric GNY theory can be relevant to address the competition between two components of Kekule valence bond solid (KVBS) (assuming the symmetry between them is explicitly broken at the microscopic level). We show that close to a quantum MCP symmetry gets enlarged to $O(2)$ that can describe transition to either one of these components that breaks the discrete $Z_2$ symmetry in the ordered phase, as well as to an $O(2)$ symmetry-breaking valence bond insulators.

2. When two order-parameters, possessing $Z_2$ and $O(2)$ symmetries, are strongly coupled, the effective field theory can be germane to describe the ultimate critical behavior of (a) an easy-plane anisotropic antiferromagnet (AF) or quantum spin-Hall insulator (QSHI), (b) competing (intra unit-cell) charge-density-wave (CDW) and complex (two-component) KVBS order or (c) competing singlet $s$-wave pairing (two component) and CDW. We demonstrate that the ultimate quantum critical behavior of such coupled order-parameter field theory is described by an $O(3)$ symmetric MCP.~\footnote{ In particular the last scenario can in principle be accessed from a microscopic attractive Hubbard-U model, which due to the underlying pseudo-spin $SU(2)$ symmetry will allow us to directly access an $O(3)$ symmetric fixed point (by causing simultaneous nucleation of CDW and $s$-wave pairing which for any strength of pure Hubbard interaction transform as an $O(3)$ vector). However, the true multi-critical nature of such fixed point can only be revealed by introducing nearest-neighbor interaction ($V_1$) in the theory, for example. Specifically, for repulsive and attractive nearest-neighbor interaction, system respectively enters into a CDW and $s$-wave paired state, while the quantum phase transition to either of these two phases (at least for $|V_1| \ll |U|$) is governed by the $O(3)$ symmetric MCP~\cite{roy-foster}.}

3. A $Z_2 \otimes O(3)$ symmetric GNY theory can be relevant to address the competition between an AF and a single component of KVBS, for example. From the leading order $\epsilon$-expansion we find that the fully stable fixed point in the critical plane possesses an $O(4)$ symmetry. In a recent numerical work such competition has been addressed using quantum Monte Carlo simulation, and findings are suggestive of an emergent $O(4)$ symmetry~\cite{sato-hohendler-assaad}, however, slightly away from the $O(4)$ symmetric itinerant MCP. Therefore, it remains an interesting future work to numerically investigate the restoration of an $O(4)$ symmetry close to this itinerant MCP, we address here. The transition between these two distinct broken symmetry phases in close proximity to the MCP is expected to be governed by the $O(4)$ symmetric MCP (see Sec.~\ref{ins-ins-mcp}), and the critical exponents determining the universality class of this transition between two ordered phases can be extracted by setting $S_1=1$ and $S_2=3$ (see Sec.~\ref{Sec:generalMCP}).

4. An $O(4)$ symmetric MCP can also result from two competing orders that break $O(2)$ symmetry in the ordered phase. Such field theoretic description is relevant to capture the competition between (a) an easy-plane AF and complex KVBS (both of them are described by two component order parameter), (b) competition between easy-plane QSHI and $s$-wave pairing, as well as (c) singlet complex (two-component) $s$-wave pairing and complex (two-component) KVBS. However, the critical properties at $O(4)$ symmetric MCP is same as the one we find from a coupled $Z_2 \otimes O(3)$ GNY model.

5. Finally, the competition between AFM and KVBS in a Dirac system is captured by an $O(2) \otimes O(3)$ symmetric GNY theory. From the leading order $\epsilon$-expansion, we demonstrate that the ultimate critical behavior is described by an $O(5)$ symmetric MCP where three-component AFM and two-component KBVS order parameters can be rotated into each other by the generators of chiral symmetry of massless Dirac fermions. The same field theory (with appropriate redefinition of spinor basis) also indicates the existence of an $O(5)$ symmetric MCP when propensity toward nucleation of QSHI (a three-component vector) and $s$-wave pairing (two-component vector) are comparable.

A comment on the connection between the largest possible symmetry restoration at an interacting MCP and the number of components of Dirac fermion is due at this stage. In graphene-like systems, where the effective low-energy theory is described in terms of an eight-component, two dimensional Dirac fermions $\mbox{max.}\left( S_1 + S_2 \right)=5$, \emph{the maximal number of mutually anti-commuting mass matrices that also anti-commute with the Dirac Hamiltonian}~\cite{okubo}. Otherwise, the possibility of emergent symmetry rests on a very simple principle: \emph{The Hermitian matrices describing two competing gapped order parameters mutually anti-commute}. Under that circumstance a \emph{composite super-vector mass} can be constructed from constituting two mass parameters and \emph{all components} of composite super-vector mass can be rotated into each other via the generators of continuous chiral symmetry of non-interacting massless Dirac fermions. Thus, ultimate quantum critical behavior is controlled by a fixed point with a maximal symmetry.

Rest of the paper is organized as follows. In the next section (Sec.~\ref{Sec:generalMCP}), we introduce an $O(S_1) \otimes O(S_2)$ symmetric GNY model and study its emergent quantum multi-critical behavior. The Sec.~\ref{ins-ins-mcp} is devoted to addressing the direct continuous quantum phase transition between $O(S_1)$ and $O(S_2)$ symmetry breaking ordered phases through the $O(S_1+S_2)$ symmetric itinerant quantum MCP. In Sec.~\ref{largen}, we discuss the anomalous scaling dimensions for local fermion bilinears at the $O(5)$ symmetric MCP within the large N framework. The summary of our central results and ramifications of emergent multi-criticality is presented in Sec.~\ref{discussion}. Additional technical details are relegated to Appendices.

\section{Coupled Gross-Neveu-Yukawa theory: Emergent Multicriticality}~\label{Sec:generalMCP}

We first introduce a \emph{master} GNY model that will allow us to establish the emergence/restoration of high symmetry near relativistic quantum MCP. Formally the effective action in $D=(d+1)$ Euclidean space-time comprised of $d$ dimensional spatial coordinates $\mathbf{r}=(x_1,..,x_d)$ and the imaginary time $\tau=x_0$, ${\mathcal S}= \int d^Dx L$ has three components, with $L=L_f+L_{Y}+L_{b}$, and $d^Dx=d^dr dx_0$. A large variety of two dimensional Dirac systems including graphene, d-wave superconductor and d-density wave can be described in terms of eight component Dirac fermions. Motivated by this we write the fermion fields as eight component Grassmann spinor $\Psi(x)$ and the conjugate spinor $\Psi^\dagger(x)$. The dynamics of massless Dirac fermions is then captured by
\begin{equation}~\label{lag:fermion}
L_f=\Psi^\dagger(x) [\partial_0-i\sum_{j=1}^{d} \; \Gamma_j \partial_j] \Psi(x).
\end{equation}
Here $\Gamma_j$s, with $j=1, \cdots, d$ are $8 \times 8$ mutually anti-commuting, Hermitian matrices, and $d \leq 7$ as there are only seven mutually anticommuting $8 \times 8$ matrices. The Yukawa coupling between massless Dirac fermions and the order-parameter fields are given by
\begin{eqnarray}~\label{lag:fermion-boson}
L_{Y} &=& g_{1} \sum^{S_1}_{j=1} \Phi_j(x) \Psi^\dagger(x) M_j \Psi(x) \nonumber \\
&+& g_{2} \sum^{S}_{j=S_1+1} \Phi_j(x) \Psi^\dagger(x) M_j \Psi(x),
\end{eqnarray}
where $M_j$s are also $8 \times 8$ mutually anticommuting, Hermitian matrices with $S=S_1+ S_2 \leq (7-d)$. Therefore, for two dimensional systems of eight component Dirac fermions, we can consider the maximal chiral symmetry group to be $O(5)$. With the same notation for three dimensional systems, we can describe $O(4)$ as the largest chiral symmetry.
The generator of (exact or emergent) chiral rotation between two mutually anti-commuting Hermitian mass matrices, namely $M_j$ and $M_k$ is $M_{jk}=-i[ M_j, M_k]/2$, which commutes with the Dirac Hamiltonian $\hat{H}_D= -i \Gamma_j \partial_j$. Therefore, $M_{jk}$'s are generators of continuous \emph{chiral} symmetry for massless Dirac fermions. For the sake of brevity we now drop the explicit dependence of bosonic and fermionic fields on $x$.

Finally, the dynamics of bosonic (order-parameter) fields is captured by $L_b=L^{1}_b+L^{2}_b+ L^{12}_b$ with
\allowdisplaybreaks[4]
\begin{eqnarray}~\label{lag:boson}
L^1_b &=& \sum^{S_1}_{j=1} \left[ \frac{1}{2} \left( \partial_\mu \Phi_j \right)^2 + m^2_1 \Phi^2_j
+\frac{\lambda_1}{4!} \left( \Phi^2_j\right)^2 \right], \nonumber \\
L^2_b &=& \sum^{S}_{j=S_1+1} \left[ \frac{1}{2} \left( \partial_\mu \Phi_j \right)^2 + m^2_2 \Phi^2_j
+\frac{\lambda_2}{4!} \left( \Phi^2_j\right)^2 \right], \nonumber \\
L^{12}_b &=& \frac{\lambda_{12}}{12} \sum^{S_1}_{j=1} \sum^{S}_{k=S_1+1} \Phi^2_j \; \Phi^2_k.
\end{eqnarray}
We note that even if we set $\lambda_{12}=0$ at the bare level, it gets generated due to two Yukawa interactions and two competing order parameter fields are coupled. Thus, we incorporate $\lambda_{12}$ in the bare action.

The form of the free fermion action implies that the engineering scaling dimension of the fermion field is $D[\Psi]=d/2$, where the scaling dimension of momentum is $D[k]=1$ and $D[\tau]=-1$. Analogously, from the free action for bosonic fields, we find that their engineering scaling dimension is $D[\Phi_j]=(d-1)/2$. This implies that scaling dimensions for the Yukawa and the four-boson couplings are respectively $D[g^2_j]=D[\lambda_j]=3-d$, while that for the mass operators $D[m^2_j]=2$. Therefore, $d=3$ is the upper critical (spatial) dimension of the theory, and in the following we will use the deviation from the upper critical dimension, namely $\epsilon=3-d$ as the expansion parameter in the renormalization group (RG) calculations. All momentum integrals are performed in $d=3-\epsilon$ dimensions, while the fermions live in two spatial dimensions, since $\Psi^\dagger i \Gamma_1\Gamma_2 \Psi$ represents a fermion mass term in two, but not in three spatial dimensions. In $d=3$ the matrix $\Gamma_3=i \Gamma_1 \Gamma_2$ is a part of the Dirac operator and neither commutes nor anticommutes with it.

\begin{figure}[t!]
\subfigure[]{
\includegraphics[width=3.95cm,height=4.0cm]{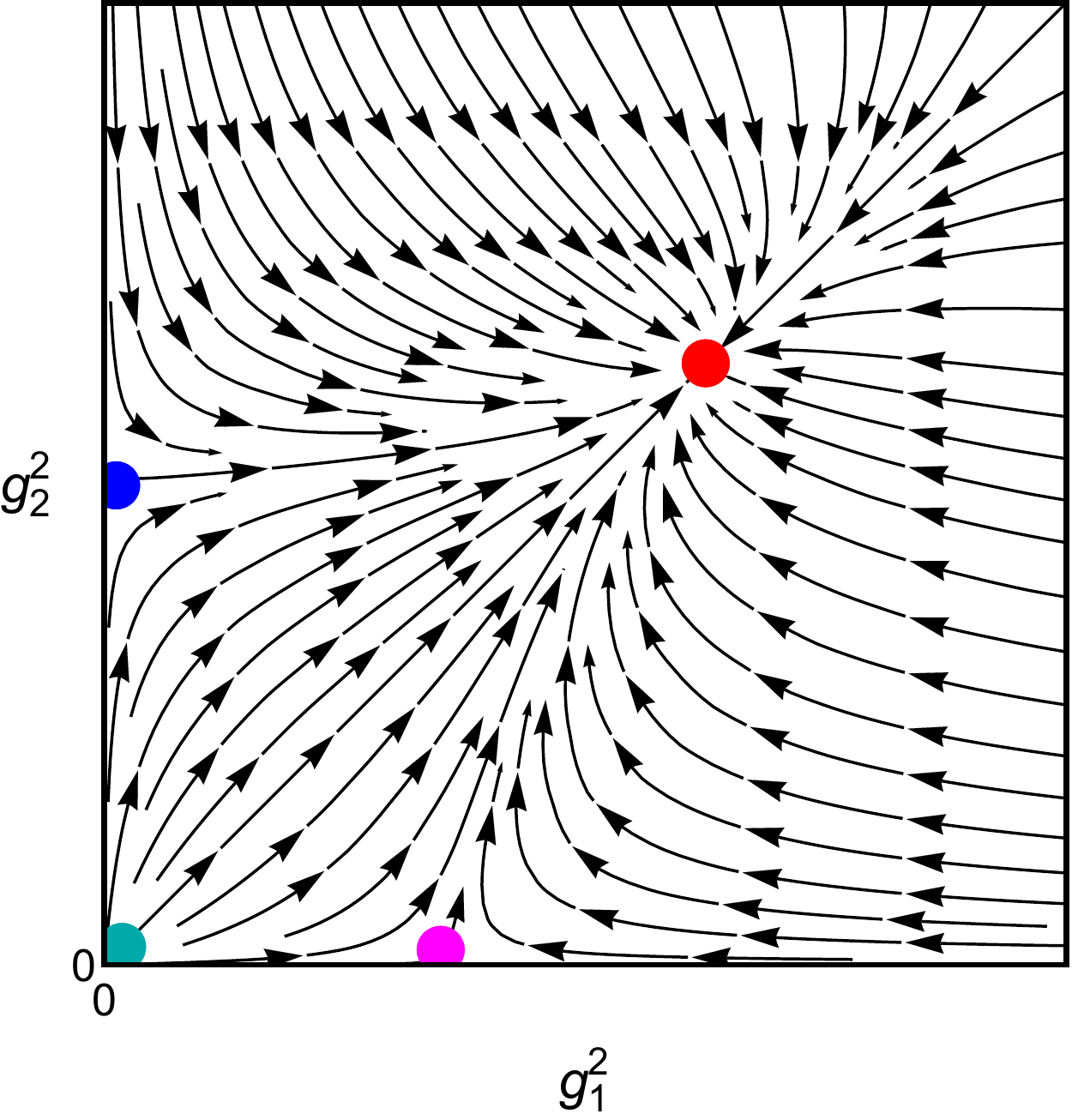}
~\label{GNY_Yukawa}}
\subfigure[]{
\includegraphics[width=3.95cm,height=4.0cm]{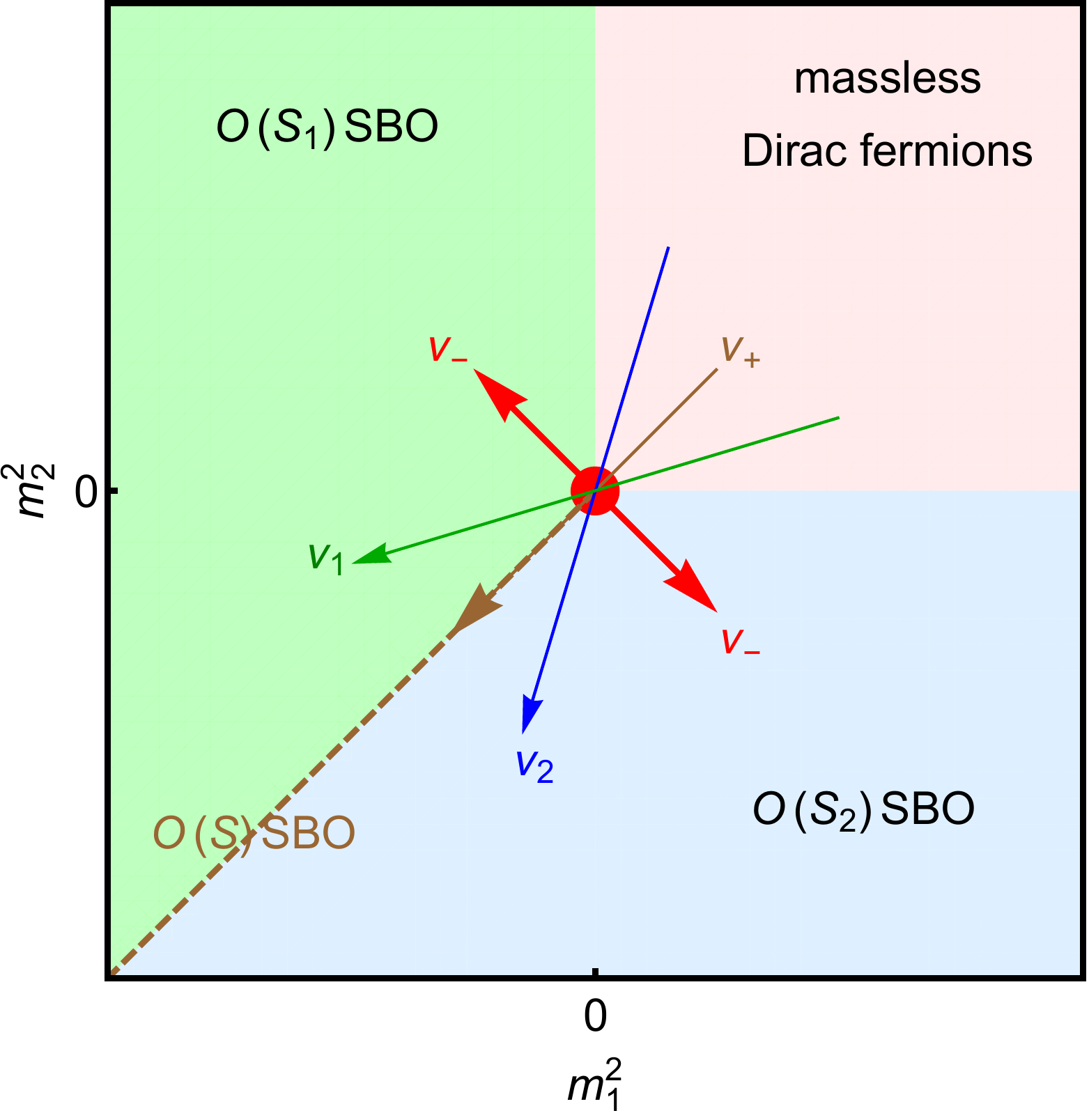}
~\label{OS_QPT}
}
\caption{ (a) A schematic renormalization group flow diagram in the plane of two competing Yukawa couplings, $g^2_1$ and $g^2_2$. All axes are measured in units of $\epsilon$, which is the deviation from the upper critical three spatial dimensions. The fixed point at $g^2_1=0=g^2_2$ (shown in dark cyan) is fully unstable. Also notice that two quantum critical points of decoupled GNY field theory located at (a) $g^2_1 \sim \epsilon$, $g^2_2=0$ (shown in magenta) and (b) $g^2_2 \sim \epsilon$, $g^2_1=0$ (shown in blue) become unstable when two fields are coupled with each other. A new quantum MCP emerges at $g^2_1=g^2_2 \equiv g^2_{\ast}  \sim \epsilon$, shown in red.
(b) A schematic phase diagram of interacting Dirac fermions residing close to $O(S_1)$ and $O(S_2)$ symmetry breaking orders (SBOs). The red dot represents the quantum MCP possessing $O(S)$ symmetry, with $S=S_1+S_2$. Note that $O(S)$ SBO can only be found for $m^2_1=m^2_2$ (along brown dashed line). Such MCP can be accessed following different routes [shown by different arrows] in the $\left(m^2_1, m^2_2\right)$ plane. Continuous transition from Dirac semimetal to (1) $O(S_1)$ SBO [green arrow], (2) $O(S_2)$ SBO [blue arrow] as well as (3) $O(S)$ SBO [brown arrow] are characterized by correlation length exponent $\nu_1$, $\nu_2$ and $\nu_+$, respectively, while that for the direct transition between $O(S_1)$ and $O(S_2)$ SBOs [red arrow] is $\nu_-$. Due to the emergent $O(S)$ symmetry at the quantum MCP [red dot], $\nu_1=\nu_2=\nu_+=\nu_-$ (see Sec.~\ref{ins-ins-mcp}). The dynamic scaling exponent is locked to $z=1$ across all transitions, manifesting emergent Lorentz symmetry at the quantum MCP.
}
\end{figure}

The RG flow equations for the above five coupling constants appearing in the imaginary time action ${\mathcal S}$ read as
\allowdisplaybreaks[4]
\begin{eqnarray}~\label{RGFlow_masterequation}
\beta_{g^2_1} &=& \epsilon g^2_1 -(2N+4-S_1) g^4_1+S_2 g^2_1 g^2_2, \nonumber \\
\beta_{g^2_1} &=& \epsilon g^2_2 -(2N+4-S_2) g^4_2+S_1 g^2_1 g^2_2, \nonumber \\
\beta_{\lambda_1} &=& \epsilon \lambda_1 - 4N g^2_1 \left( \lambda_1 -6 g^2_1 \right) -\frac{1}{6} \left[ \lambda^2_1 (8+S_1) + S_2 \lambda^2_{12} \right], \nonumber \\
\beta_{\lambda_2} &=& \epsilon \lambda_2 - 4N g^2_2 \left( \lambda_2 -6 g^2_2 \right) -\frac{1}{6} \left[ \lambda^2_2 (8+S_2) + S_1 \lambda^2_{12} \right], \nonumber \\
\beta_{\lambda_{12}} &=& \epsilon \lambda_{12} -2 N \left( g^2_1 +g^2_2 \right) \lambda_{12} + 24 N g^2_1 g^2_2 \nonumber \\
&-& \frac{\lambda_{12}}{6} \left[ (2+S_1) \lambda_1 + (2+S_2) \lambda_2 + 4 \lambda_{12} \right],
\end{eqnarray}
to the leading order in $\epsilon$-expansion, in terms of dimensionless coupling constants $X q^{-\epsilon}/(8 \pi^2) \to X$, where $X=g^2_1, g^2_2, \lambda_{1}, \lambda_2, \lambda_{12}$, and $q$ is the momentum scale defining infrared RG flow equation for a coupling $X$, with $\beta_X=-dX/d\ln q$. In the above flow equation $N$ counts the number of 4-component Dirac fermions, and thus for a graphene-like system $N=2$. Note the flow equations for Yukawa couplings $g^2_1$ and $g^2_2$ are decoupled from the flow equations of four-boson couplings ($\lambda_1$, $\lambda_2$ and $\lambda_{12}$), so we first analyze $\beta_{g^2_1}$ and $\beta_{g^2_2}$. These two coupled flow equations supports only one stable fixed point located at
\begin{equation}~\label{FP:locationgeneral_Yukawa}
g^2_1=g^2_2=\frac{\epsilon}{2 N+4-S}=g^2_\ast \; (\mbox{say}).
\end{equation}
This scenario is schematically depicted in Fig.~\ref{GNY_Yukawa}. Two stable decoupled fixed points located at
\begin{equation}
\left( g^2_1, g^2_2 \right)= \left( \frac{\epsilon}{2 N + 4-S_1}, 0 \right) \:\: \mbox{and} \:\:
\left( 0, \frac{\epsilon}{2 N + 4-S_2} \right)
\end{equation}
in a pure $O(S_1)$ and $O(S_2)$ symmetric GNY theory becomes unstable toward a new attractive fixed point due to the coupling between these two sectors.

Upon substituting the fixed point values of two Yukawa couplings in the flow equations for $\lambda_1$, $\lambda_2$ and $\lambda_{12}$, we find that three coupled flow equations support only two fixed points at $\lambda_1=\lambda_2=\lambda_{12}=\lambda^\pm_\ast$, where
\begin{eqnarray}~\label{FP:locationgeneral_boson}
\lambda^{\pm}_\ast &=& \frac{3}{(8+S) (2N+4-S)} \bigg[  \left(4-2N-S \right) \nonumber \\
&\pm&  \sqrt{4 N^2 + (S-4)^2 + 4 N (28 + 5S)}   \bigg] \epsilon.
\end{eqnarray}
Only $\lambda^+_\ast>0$ and therefore lies in the physical coupling regime. Upon setting $g^2_1=g^2_2=g^2_\ast$ and $\lambda_1=\lambda_2=\lambda_{12}=\lambda^+_\ast$ the imaginary time action ${\mathcal S}$ assumes an $O(S)$ symmetric form in the critical plane $m^2_1=m^2_2=0$.
Before further exploring the quantum multi-critical behavior, it is worth pausing to discuss the topology of the RG flow in an $O(S)$ symmetric critical plane $\left(g^2, \lambda \right)$ (with $m^2_1=m^2_2=m^2=0$), where $g^2=g^2_1=g^2_2$ and $\lambda=\lambda_1=\lambda_2=\lambda_{12}$, shown in Fig.~\ref{GNY_Final}.

Coupled flow equation of $g^2$ and $\lambda$ altogether support four fixed points in the $m^2=0$ plane. The one at $\left( g^2, \lambda \right)=(0,0)$ is fully unstable (dark cyan circle). The Wilson-Fished (WF) fixed point (shown in blue) located at $\left( g^2, \lambda \right)=(0,\frac{6}{8+S}\epsilon)$ becomes unstable due to boson-fermion Yukawa coupling. The bi-critical point (magenta circle), located at $\left( g^2, \lambda \right)=(g^2_\ast,\lambda^-_\ast)$ lies in the unphysical regime since $\lambda^-_\ast<0$. The fully stable fixed point (red circle) is located at $\left( g^2, \lambda \right)=(g^2_\ast,\lambda^+_\ast)$ and possesses only one unstable direction along $m^2$ (bosonic mass). Note that the bosonic mass plays the role of a tuning parameter for the quantum phase transition between  disordered (semimetallic) and ordered phases, respectively realized for $m^2>0$ and $m^2<0$. Now we return to the discussion on multi-critical behavior of $O(S_1) \otimes O(S_2)$ theory.

The flow equations of two mass parameters read as
\begin{eqnarray}~\label{RG:masses}
\beta_{m^2_1} &=& m^2_1 \left( 2 - 2 N g^2_1 -\frac{1}{6} \left[ (2+S_1) \lambda_1 + S_2 \lambda_{12} \right] \right), \nonumber \\
\beta_{m^2_2} &=& m^2_2 \left( 2 - 2 N g^2_2 -\frac{1}{6} \left[ (2+S_2) \lambda_2 + S_1 \lambda_{12} \right] \right).
\end{eqnarray}
Since each mass parameter tunes the quantum phase transition of massless Dirac fermions into an ordered phase, we are dealing with a MCP, see Fig.~\ref{OS_QPT}. The correlation length exponents associated with the quantum phase transitions into the ordered phase breaking $O(S_1)$ symmetry (with $\langle \Phi_j \rangle \neq 0$ for $j=1, \cdots, S_1$) and $O(S_2)$ symmetry (with $\langle \Phi_j \rangle \neq 0$ for $j=S_1+1, \cdots, S$) are respectively given by
\begin{eqnarray}
\nu_1 &=& \frac{1}{2} + \frac{N}{2} g^2_1 + \frac{1}{24} \left[ (2+S_1) \lambda_1 + S_2 \lambda_{12} \right], \nonumber \\
\nu_2 &=& \frac{1}{2} + \frac{N}{2} g^2_2 + \frac{1}{24} \left[ (2+S_2) \lambda_2 + S_1 \lambda_{12} \right],
\end{eqnarray}
since $\beta_{m^2_1}=m^2_1 \nu^{-1}_1$ and $\beta_{m^2_2}=m^2_2 \nu^{-1}_2$. Notice that at $O(S)$ symmetric MCP $\nu_1=\nu_2=\nu^{(S)}$ (say), with
\begin{equation}~\label{CLE:final}
\nu^{(S)}=\frac{1}{2} + \frac{N}{2} g^2_\ast + \frac{2+S}{24} \lambda^+_\ast,
\end{equation}
with $g^2_\ast$ and $\lambda^{+}_\ast$ are resectively given in Eq.~(\ref{FP:locationgeneral_Yukawa}) and Eq.~(\ref{FP:locationgeneral_boson}). \emph{Therefore, correlation length exponents associated with the quantum phase transition from a Dirac semimetal to an ordered phase that spontaneously breaks either $O(S_1)$ or $O(S_2)$ symmetry (respectively given by $\nu_1$ and $\nu_2$), as well as $O(S_1+S_2)$ symmetry (given by $\nu^{(S)}$), through the quantum MCP are equal and given by $\nu^{(S)}$, quoted in Eq.~(\ref{CLE:final})}. With increasing chiral symmetry at the GNY MCP $\nu^{(S)}$ increases monotonically.

To the leading order in $\epsilon$-expansion the bosonic anomalous dimension at a $O(S)$ symmetric MCP is given by
\begin{equation}~\label{anomalousdim:Bos}
\eta^{(S)}_b = 2N g^2_\ast =\frac{2N}{2N+4-S} \; \epsilon.
\end{equation}
Therefore, with increasing number of the order-parameter components $S$, the bosonic anomalous dimension increases monotonically and acquires largest value at the $O(5)$ symmetric MCP for $N=2$.
We also note that the fermionic anomalous dimension
\begin{equation}~\label{anomalousdim:Fer}
\eta^{(S)}_\Psi= \frac{S}{2}\; g^2_\ast= \frac{S}{2 \left( 2N+4 -S\right)} \; \epsilon,
\end{equation}
increases monotonically as the symmetry enlarges at the quantum MCP.

These exponents determine scaling of various physical observables. For example, fermionic Greens function in the close proximity to an $O(S)$ symmetric MCP scales as
\begin{equation}
G^{-1}_f \sim \left( \omega^2 + k^2 \right)^{\left(1-\eta_\Psi\right)/2}.
\end{equation}
Hence, the residue of fermionic quasiparticle pole vanishes as
\begin{equation}
Z_\Psi \sim \left( m \right)^{\frac{S}{4 (2N+4-S)} \; \epsilon}
\end{equation}
Therefore, with increasing symmetry gapless fermionic excitations lose their sharpness more rapidly as the MCP is approached from the semimetallic side of the transition, and at the MCP the notion of sharp quasiparticles becomes moot. As the MCP is approached from the ordered side the fermionic and bosonic masses vanish smoothly, however, with the following universal ratio
\begin{equation}
\left( \frac{m_b}{m_f} \right)^2=\frac{\lambda^{+}_\ast}{3g^2_\ast}.
\end{equation}
To the extreme close proximity of the GNY MCP the ratio of specific heat in the disordered (Dirac semimetal) and ordered side is also a universal quantity, and as  $T \to 0$ it is given by
\begin{equation}
\lim_{T \to 0} \frac{C_{dis}}{C_{order}} = \frac{4N}{N_G} \left(1-2^{-d} \right)= \frac{4N}{S-1} \left(1-2^{-d} \right),
\end{equation}
for $S>1$, where $N_G=S-1$ is the number of Goldstone modes in the ordered phase. We here tacitly assumed that fermionic and bosonic excitations are characterized by same velocity, which is a direct manifestation of emergent Lorentz symmetry at the GNY critical point.

We can also estimate the dc conductivity due to inelastic scattering at the $O(S_1+S_2)$ symmetric MCP. The dc conductivity would follow the form of Drude formula $\sigma \sim ne^2 \tau/m$ with $m=k_BT/v^2$ as the effective mass, and $n \sim N \; (k_BT)^2/(\hbar v)^2$ being the density of thermally excited carriers (equal number of particles and holes). As long as $T/T_0 \ll \delta^{1/\epsilon}$ where $\delta=|g^2-g^2_\ast|$ measures the deviation of the Yukawa coupling from its fixed point value, the inelastic scattering rate is approximately given by $\hbar/\tau \sim (S_1+S_2) \; g^4_\ast \; (k_B T)$. Therefore, the dc conductivity at (or in the vicinity of) the MCP is estimated to be
\begin{equation}
\sigma \sim \frac{e^2}{\hbar} \; \frac{N}{S_1+S_2} \; \frac{1}{g^4_\ast} \; \left(\frac{k_B T}{\hbar v}\right)^{(d-2)}.
\end{equation}
Since $g^2_\ast \sim (3-d)/N$, the MCP at d=2 has a dc conductivity $\sigma \sim \frac{e^2 N^3}{\hbar (S_1+S_2)}$. By contrast the conductivity at the lower symmetric $O(S_1)$ fixed point would be $\sigma \sim \frac{e^2 N^3}{\hbar (S_1)}$. \emph{Therefore, the dc resistivity is enhanced at the quantum MCP in comparison to the lower symmetric critical points, which is in agreement with the stronger non-Fermi liquid properties at the MCP.} A direct computation of dc conductivity at the quantum MCP is left for a future investigation (see however Ref.~\cite{lars}). In comparison to Ref.~\cite{lars}, reporting $\sigma \sim N^2$, we find $\sigma \sim N^3$. The additional factor of $N$ in the expression of the dc conductivity in our analysis arises from the dependence of the carrier density $n \sim N$ on the fermion flavor number. We also note that the optical conductivity (in the collisionless regime) gets reduced at the interacting MCP, in comparison to its counterpart in noninteracting Dirac liquid~\cite{roy-juricic-OC}


\section{Direct Transition between two ordered phases}~\label{ins-ins-mcp}

Previously we have shown that quantum phase transition of massless Dirac fermions into an $O(S_1)$, $O(S_2)$, $O(S)$ symmetry breaking phases share the same universality class when all of them are controlled by an $O(S)$ symmetric MCP. Namely, the bosonic ($\eta^{(S)}_b$) and fermionic ($\eta^{(S)}_\Psi$) anomalous dimensions, and the correlation length exponent ($\nu^{(S)}$) across all three transitions are the same. \emph{Furthermore, the $O(S)$ symmetric quantum MCP also controls the direct quantum phase transition between $O(S_1)$ and $O(S_2)$ symmetry breaking ordered phases}, see Fig.~\ref{OS_QPT}. In this section we address the critical properties of such a direct transition between two competing ordered phases.

We now introduce a new set of variables
\begin{eqnarray}
g^2_{\pm} = \frac{g^2_1 \pm g^2_2}{2}, \:
\lambda_{\pm} =\frac{\lambda_1 \pm \lambda_2}{2}, \:
m^2_{\pm}= \frac{m^2_1 \pm m^2_2}{2}.
\end{eqnarray}
In this notation $m^2_+$ controls the continuous transition from Dirac semimetal to an $O(S)$ symmetry breaking ordered phase. On the other hand, $m^2_-$ controls the transition between an $O(S_1)$ symmetry breaking ordered phase (for $m^2_+<0$ and $m^2_-<0$) to an $O(S_2)$ symmetry breaking ordered phase (for $m^2_+<0$ and $m^2_->0$). The RG flow equation of $m^2_-$ can then be obtained from Eq.~(\ref{RG:masses}), yielding
\begin{widetext}
\allowdisplaybreaks[4]
\begin{eqnarray}
\beta_{m^2_-} &=& 2 \left[ m^2_- - N \left( g^2_+ m^2_- + g^2_- m^2_+ \right)\right]-\frac{1}{12} \bigg[ (S+4) \bigg\{ m^2_- \left(\lambda_+ + \lambda_{12} \right) + m^2_+ \lambda_- \bigg\} -4 m^2_- \lambda_{12} \nonumber \\
&+& \left( S_1-S_2\right) \bigg\{ m^2_- \lambda_- + m^2_+ \left( \lambda_+ -\lambda_{12} \right) \bigg\} \bigg].
\end{eqnarray}
\end{widetext}
In the extreme close proximity to the $O(S)$ symmetric quantum MCP we set set $g^2_-=0$, $\lambda_-=0$ and $\lambda_{12}=\lambda_+$. For the continuous phase transition between two ordered phases we follow the trajectory $m^2_+=0$ and set $m^2_+=0$. The above flow equation then simplifies to
\begin{equation}
\beta_{m^2_-}= m^2_- \left[2-2N g^2_+ -\frac{S+2}{6} \lambda_+  \right].
\end{equation}
Therefore, the correlation length exponent associated with the direct continuous transition between $O(S_1)$ and $O(S_2)$ symmetry breaking ordered phases, given by $\nu_-$ and defined as $\beta_{m^2_-}=m^2_- \nu^{-1}_-$, through the $O(S)$ symmetric itinerant quantum MCP is given by
\begin{equation}
\nu_- = \nu^{(S)},
\end{equation}
which is same as the ones for the continuous quantum phase transition from a Dirac semimetal to $O(S_{1})$ or $O(S_{2})$ or $O(S)$ symmetry breaking ordered phase. This situation is qualitatively displayed in Fig.~\ref{OS_QPT}.

Irrelevance of $g^2_-$ and $\lambda_-$ in the flow equation of $m^2_-$ can be demonstrated in the following way. The leading order correction to $g^2_-$ is given by
\begin{eqnarray}
\beta_{g^2_-}&=& g^2_- \left[ \epsilon- (4 N +8-S) g^2_+ \right] + {\mathcal O} (g^4_-) \nonumber \\
             &=& -\epsilon g^2_- \: \left[ \frac{2 N +4}{2N+4-S} \right] + {\mathcal O} (g^4_-),
\end{eqnarray}
confirming that $g^2_-$ is an irrelevant perturbation close the $O(S)$ symmetric MCP. On the other hand, irrelevance of $O(S)$ symmetry breaking four-boson interaction can be demonstrated by rewriting the interacting part of the bosonic action as
\begin{eqnarray}
L_b = \frac{ \lambda}{4!} \sum^{S}_{j=1} \left( \Phi^2_j\right)^2
+ \frac{\delta\lambda_{12}}{12} \sum^{S_1}_{j=1} \sum^{S_1+S_2}_{k=S_1+1} \Phi^2_j \Phi^2_k.
\end{eqnarray}
with $\lambda=\lambda_1=\lambda_2=\lambda_{12}$. The coupling constant $\delta \lambda_{12}$ can be considered as an $O(S)$ symmetry breaking term. The leading order correction to $\delta \lambda_{12}$ is given by
\begin{eqnarray}
\beta_{\delta\lambda_{12}} = \delta \lambda_{12} \left[ \epsilon-4 N g^2_+ -\frac{S+4}{6} \; \lambda \right] + {\mathcal O} (g^4_+,g^4_-, \delta \lambda^2_{12}), \nonumber \\
\end{eqnarray}
Substituting $g^2_+=g^2_\ast$ and $\lambda=\lambda^{+}_\ast$, readers can immediately confirm that $\delta \lambda_{12}$ is also an irrelevant parameter close to the $O(S)$ symmetric quantum MCP. Irrelevance of two $O(S)$ symmetry-breaking perturbations (namely $g_-$ and $\delta\lambda_{12}$) suggests that the $O(S)$ symmetric MCP is accompanied by a finite domain of attraction. Consequently, as long as the transition between two distinct ordered phases, breaking $O(S_1)$ and $O(S_2)$ symmetries such that $S=S_1+S_2$, takes place sufficiently close to the $O(S)$ symmetric MCP, the universality class of such order-order transition is characterized by the one at such MCP, with exponents being $\nu^{(S)}$ [see Eq.~(\ref{CLE:final})], $\eta^{(S)}_b$ [see Eq.~(\ref {anomalousdim:Bos})] and $\eta^{(S)}_\Psi$ [see Eq.~(\ref{anomalousdim:Fer})]. In conjunction with the irrelevance of the $O(S)$ symmetry breaking perturbations [namely $g_-$ and $\delta\lambda_{12}$], the fact that both $O(S_1)$ and $O(S_2)$ symmetry breaking phases always occupy a finite regime in the coupling constant space [see Fig.~\ref{OS_QPT}], ensures that direct order-order continuous transition can be controlled by the $O(S)$ symmetric itinerant MCP as long as this transition takes place sufficiently close to the MCP. In other words, \emph{the order-order transition does not need to be fine-tuned directly through the MCP to realize the associated universality class of the transition}.

Finally we point out that deep inside the ordered phases (sufficiently far from the itinerant MCP), the direct transition between two ordered phases can be controlled by an entirely different critical point. However, such a critical point cannot be accessed within the framework of GNY theory. Nevertheless, our present discussion provides unprecedented insights into the direct continuous quantum phase transition between two distinct broken symmetry phases through an itinerant quantum MCP.
 We note that the numerical work~\cite{sato-hohendler-assaad} found the appearance of an $O(4)$ symmetry slightly away from the MCP. It will therefore be extremely interesting to analyze the universality class of the continuous transition between $O(3)$ and $Z_2$ symmetry breaking phases close to the itinerant MCP and test the validity of our prediction.
In the following section, we provide further insight into the nature of itinerant MCP by performing a large $N$ (number of eight component Dirac fermions) analysis in $d=2$ dimensions. We will also compute the anomalous scaling dimensions of all local fermion bilinears in the particle-hole channel, to gain a more comprehensive understanding of critical correlation functions.


\section{Large $N$ description}\label{largen}

For a large $N$ analysis, we first scale the Yukawa couplings as $g_j \to g_j/\sqrt{N}$, and the fermion bubble contribution to the  boson propagator becomes an $O(1)$ quantity. Therefore in the $N \to \infty$ limit, the propagator for order parameter fields at the MCP would be given by $D(q)=2/q$, where $q=\sqrt{q^2_0+\mathbf{q}^2}$. The $O(1/N)$ correction to the fermion fields is obtained to be $\eta_\psi=S/(3N \pi^2)$. These results are in general consistent with taking the large $N$ limit of the exponents $\eta_b$ and $\eta_\psi$ obtained from the GNY theory. Now we will calculate the anomalous scaling dimensions of the fermion bilinears for the O(5) MCP.

For convenience we adopt the following notations for the $64$ eight dimensional matrices. For the order parameter fields we choose $M_j=\gamma_j \otimes \eta_3$, where $\gamma_j$'s are five mutually anticommuting Hermitian matrices, and $\eta_3$ is the diagonal Pauli matrix. Then the additional two anticommuting $8 \times 8$ matrices for the Dirac propagator are given by $\mathbb{1} \otimes \eta_1$ and $\mathbb{1} \otimes \eta_2$, where $\mathbb{1}$ is the $4\times 4$ identity matrix and $\eta_{1,2}$ are the two off-diagonal Pauli matrices. Therefore, the ten generators for the O(5) chiral symmetry are given by $M_{jk}=\gamma_{jk} \otimes \eta_0$ with $\eta_0$ being the $2 \times 2$ identity matrix, and $\gamma_{jk}=[\gamma_j,\gamma_k]/(2i)$. Now all other $8 \times 8$ matrices can be written down as $\gamma_j \otimes \eta_0$, $\gamma_j \otimes \eta_1$, $\gamma_j \otimes \eta_2$, $\mathbb{1} \otimes \eta_0$, $\mathbb{1} \otimes \eta_3$, $\gamma_{jk} \otimes \eta_3$, $\gamma_{jk} \otimes \eta_1$, $\gamma_{jk} \otimes \eta_2$. In terms of these matrices we can construct all the local fermion bilinears ($\Psi^\dagger \mathcal{O} \Psi$) in the particle-hole channel, and they describe general form of charge, spin, valley density and current operators.

At the free fermion fixed point, any local fermion bilinear $\Psi^\dagger \mathcal{O} \Psi$ has the scaling dimension $\Delta=d$. Therefore, the generalized four-fermion or density-density correlation function $\langle \Psi^\dagger (x)\mathcal{O} \Psi(x) \Psi^\dagger (x^\prime)\mathcal{O} \Psi (x^\prime) \rangle$ decays as $1/|x-x^\prime|^{2\Delta}=1/|x-x^\prime|^4$. If we denote a source field coupled to $\Psi^\dagger \mathcal{O} \Psi$ as $\lambda_{\mathcal{O}}$,
\begin{eqnarray}
\frac{d\lambda_{\mathcal{O}}}{d\ell}&=&\lambda_{\mathcal{O}}(1+\eta_{\mathcal{O}}-\eta_\psi), \\
\Delta_{\mathcal{O}}&=&D-1-\eta_{\mathcal{O}}+\eta_\psi,
\end{eqnarray} 
where we have to determine $\eta_{\mathcal{O}}$ from a vertex involving $\mathcal{O}$, dressed by the boson propagator and $\ell$ is the logarithm of the RG scale. Since $\mathcal{O}_{1}=\mathbb{1} \otimes (\eta_0, \eta_1, \eta_2)$ correspond to charge and current operators for total number, they do not acquire any anomalous scaling dimension from the order parameter couplings (i.e., $\eta_{1,\mu}=\eta_\psi$). By contrast, the density and current operators for the emergent O(5)symmetry (including $\pi$ operators~\footnote{
Let us denote the spin-triplet antiferromagnetic bilinears by $\Psi^\dagger \gamma_j \otimes \eta_3 \Psi$ with $j=1,2,3$. The spin-singlet valence bond bilinears are then described by $\Psi^\dagger \gamma_j \otimes \eta_3 \Psi$ with $j=4,5$. Naturally $\Psi^\dagger \gamma_{12} \Psi$, $\Psi^\dagger \gamma_{23} \Psi$, $\Psi^\dagger \gamma_{31} \Psi$ cause rotation among the components of antiferromagnetic order, while $\Psi^\dagger \gamma_{45} \Psi$ is responsible for O(2) rotation between two components of VBS order. The following six operators $\Psi^\dagger \gamma_{14} \Psi$, $\Psi^\dagger \gamma_{15} \Psi$, $\Psi^\dagger \gamma_{24} \Psi$, $\Psi^\dagger \gamma_{25} \Psi$, $\Psi^\dagger \gamma_{34} \Psi$, $\Psi^\dagger \gamma_{35} \Psi$ cause general chiral rotation between triplet and singlet order parameters. In the context of $SO(5)$ theory of high temperature superconductivity such operators, rotating between the antiferromagnetic and d-wave superconducting order-parameters, were named as $\pi$ operators~\cite{ZhangDemler}.}) are captured by $\mathcal{O}_{2}=\gamma_{jk} \otimes (\eta_0, \eta_1, \eta_2)$, and  we find that
\begin{equation}
\Delta_{2}=2+ \frac{4}{3N\pi^2}.
\end{equation} Therefore, these operators display faster algebraic decay (or shorter range power law correlations), in comparison with free fermion fixed point. We also note that the generators of lower $O(S_1)$ or $O(S_2)$ symmetry (emergent conserved quantities of an interacting fixed point) would also exhibit the same scaling dimension at the respective, decoupled quantum critical points. The scaling dimensions for $\mathcal{O}_3=\mathbb{1} \otimes \eta_3$, $\mathcal{O}_4=\gamma_{jk} \otimes \eta_3$, $\mathcal{O}_{5}=\gamma_j \otimes (\eta_0, \eta_1, \eta_2)$, are respectively given by
\begin{eqnarray}
\Delta_3&=&d+\frac{4S}{3N \pi^2}=2+\frac{20}{3N \pi^2}, \\
 \Delta_4&=&d+\frac{2(S-2)}{3N \pi^2}=2+\frac{10}{3N\pi^2}, \\
\Delta_{5}&=&d+\frac{2(S-1)}{3N\pi^2}=2+\frac{8}{3N \pi^2}.
\end{eqnarray} Therefore, all four fermion correlation functions at the MCP (other than the total number or current operators) exhibit faster power law decays at large distance and time. This is consistent with the fact that there are no other anticommuting matrices or parameters which can possess slower decay or potential instability of the MCP. By contrast, the lower symmetric $O(S_1)$ critical point would show slower decay or longer ranged correlations for the $O(S_2)$ anticommuting operators. Moreover, the presence of anomalous dimension $\Delta \neq 2$ indicates that the dynamic structure factor for all possible generalized scattering experiments would show a continuum of excitations in accordance with the lack of quasiparticle poles for both fermionic and bosonic excitations. We would also like to mention that as a consequence of the non-Gaussian nature of the MCP, all dynamic correlations functions at finite temperature would exhibit $\omega/T$ scaling.


\section{Discussions}~\label{discussion}

To summarize, we here show that when massless Dirac fermions residing in two-spatial dimensions acquire comparable propensity toward formation to two distinct broken symmetry phases, that in the ordered phase break $O(S_1)$ and $O(S_2)$ symmetries but can be chirally rotated into each other, the ultimate quantum critical behavior is governed by an $O(S)$ symmetric multi-critical point, where $S=S_1+S_2$. For a graphene-like system, accommodating eight-component massless Dirac fermions $\mbox{max.}(S)=5$. Such quantum multi-critical point can control the continuous phase transition from Dirac semimetal to either $O(S_1)$ or $O(S_2)$ or $O(S)$ symmetry breaking ordered phases, as well as between two distinct broken symmetry phases. All the transitions are characterized by identical correlation length exponent $\nu^{(S)}$ (manifesting restoration of $O(S)$ symmetry at the quantum multi-critical point) and dynamic scaling exponent $z=1$ (stemming from emergent Lorentz symmetry).

Even though we generically find the existence of $O(S)$ symmetric multi-critical point from a leading order $\epsilon$-expansion about the upper critical three spatial dimensions with $\epsilon=3-d$, the restoration of symmetry among dual mass orders may be non-perturbative in nature for the following reason. \emph{It is believed that quantum criticality in a coupled field theory is ultimately governed by a fixed point with largest bosonic anmalous dimension $\eta_b$}~\cite{zinn-justin-book-2, zinn-justin-vicari}. Such an outcome is expected to hold for pure bosonic as well as boson-fermion coupled Gross-Neveu-Yukawa models. If we denote the bosonic anomalous dimensions at $O(S_1)$ and $O(S_2)$ symmetric critical points in decoupled theory respectively as $\eta^{(S_1)}_b$ and $\eta^{(S_2)}_b$, we always find that
\begin{equation}
\eta^{(S)}_b>\eta^{(S_1)}_b, \eta^{(S_2)}_b
\end{equation}
for $S=S_1+S_2$, at least to the leading order in $\epsilon$-expansion [see Eq.~(\ref{anomalousdim:Bos})]. From the present discussion we can further strengthen the stability criterion for a multi-critical point with enlarged symmetry as follows: \emph{The ultimate quantum multi-critical behavior should be controlled by a fixed point displaying largest fermionic anomalous dimension and largest correlation length exponent.} Although, we did not find any violation of this rule so far, this observation is purely based on perturbative analysis, and we leave it at the state of a \emph{conjecture}. Nonetheless, in Appendix~\ref{Sec:largestsymm} we show that the ultimate quantum multi-criticality of a coupled Gross-Neveu-Yukawa model that can potentially support $O(S)$ symmetric fixed points, with $S=1,2,3,4$, is controlled by an $O(4)$ symmetric fixed point, in agreement with our proposed conjecture.

\begin{figure*}[t!]
\subfigure[]{
\includegraphics[width=4.1cm,height=3.35cm]{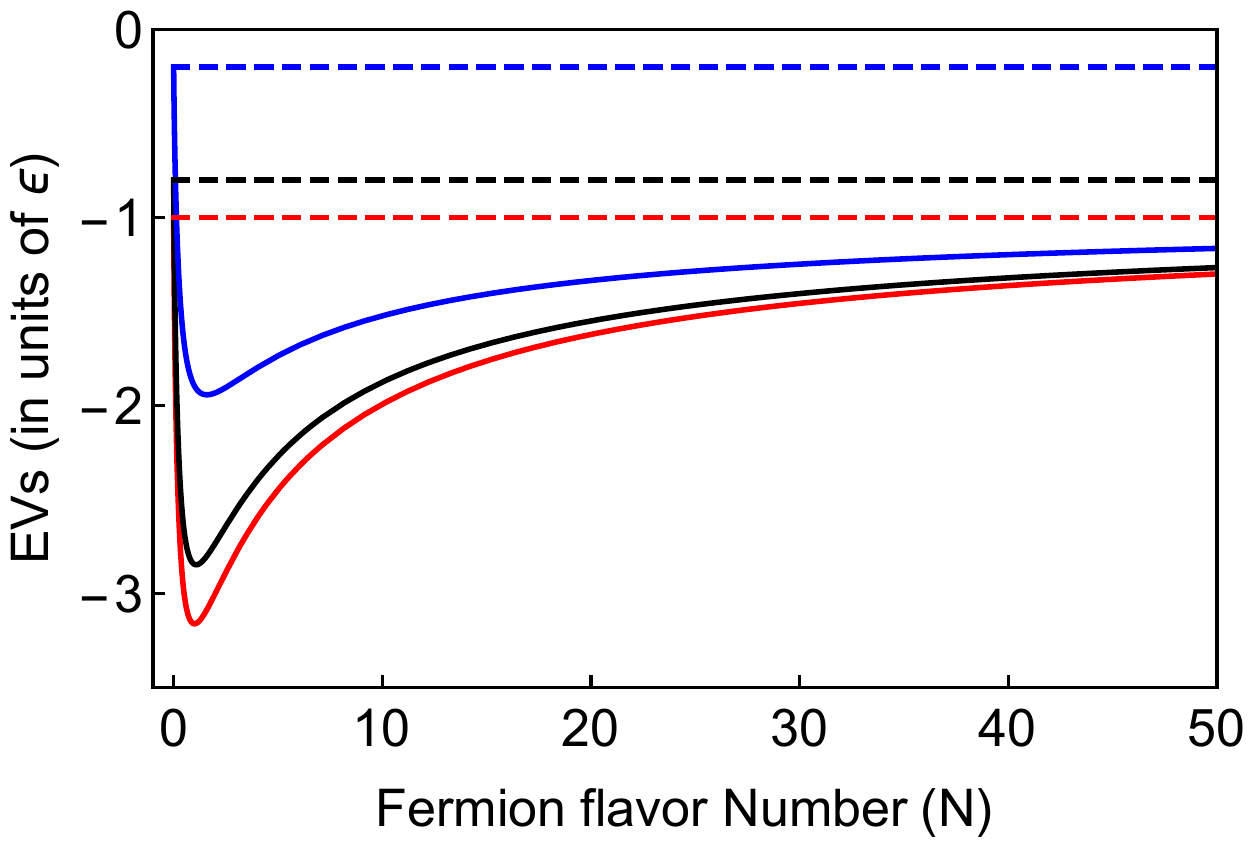}
\label{EV_O2}
}
\subfigure[]{
\includegraphics[width=4.1cm,height=3.25cm]{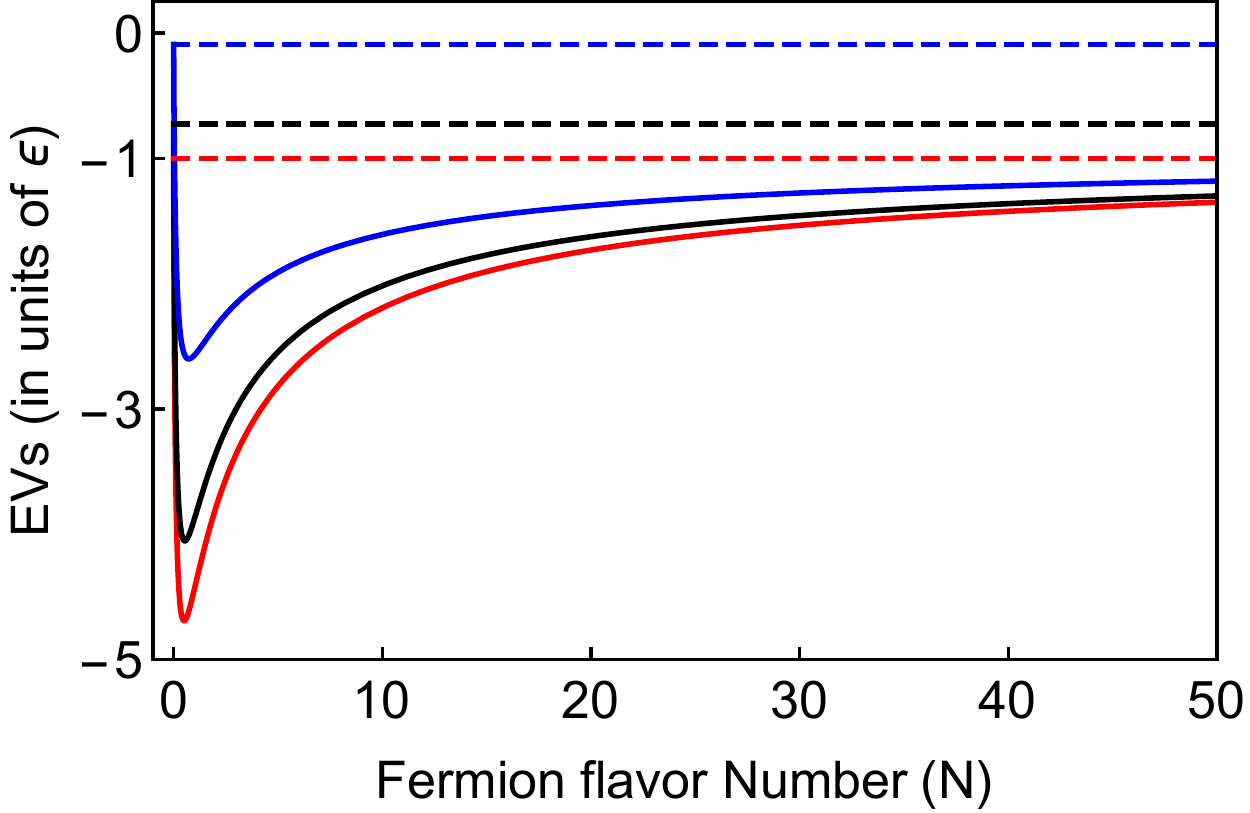}
\label{EV_O3}
}
\subfigure[]{
\includegraphics[width=4.1cm,height=3.25cm]{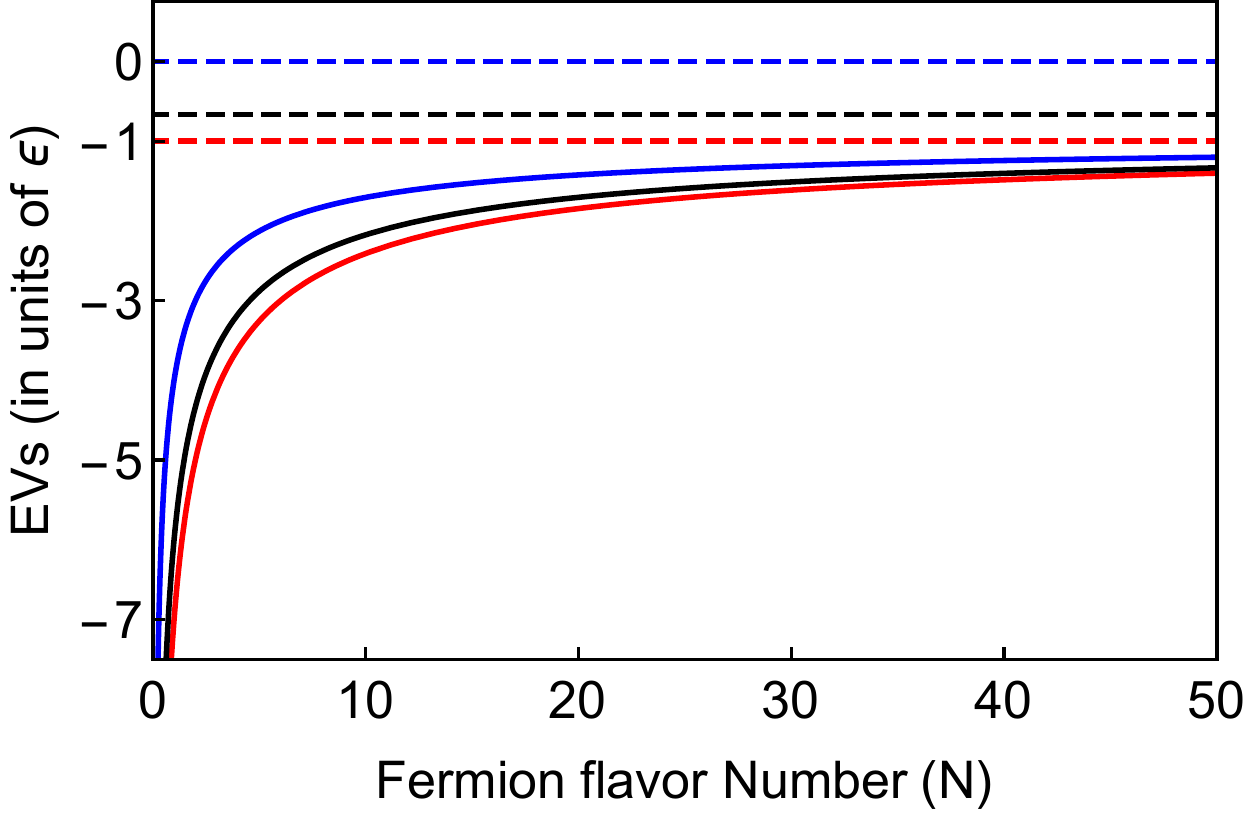}
\label{EV_O4}
}
\subfigure[]{
\includegraphics[width=4.1cm,height=3.25cm]{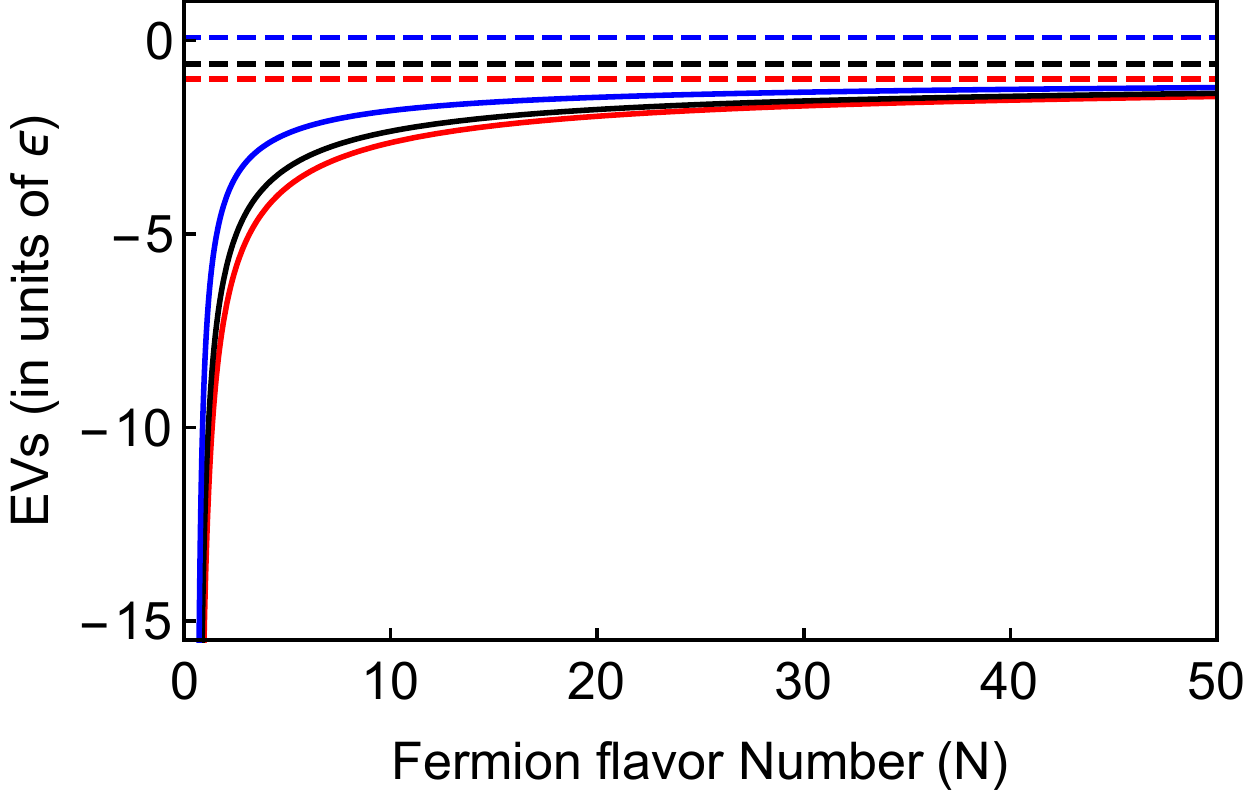}
\label{EV_O5}
}
\caption{ Scaling of three eigenvalues (EVs), measured in units of $\epsilon$, of the stability matrix in the three dimensional coupling constant space spanned by four-boson interactions ($\lambda_1$, $\lambda_2$ and $\lambda_{12}$) near (a) $O(2)$, (b) $O(3)$, (c) $O(4)$ and (d) $O(5)$ symmetric multi-critical points (MCPs) with fermion flavor number ($N$) in a pure bosonic model (dashed lines), and when bosonic and fermionic degrees of freedom are coupled with each other via Yukawa coupling (solid lines). Note in $Z_2 \otimes Z_2$ and $Z_2 \otimes O(2)$ symmetric pure bosonic relativisitc theories, the leading order $\epsilon$-expansion indicates the emergence of fully stable $O(2)$ and $O(3)$ symmetric MCP, respectively, and thus as $N \to 0$ we recover the EVs of the stability matrix in pure bosonic theory [see also Table~\ref{z2z2_bosonic} and Table~\ref{z2o2_bosonic}, respectively]. By contrast, there is no analog of $O(4)$ [resulting from $Z_2 \otimes O(3)$ or $O(2) \otimes O(2)$ theory] or $O(5)$ [resulting from $O(2) \otimes O(3)$ theory] symmetric MCPs in pure bosonic theory, and the EVs as $N \to 0$ are not connected with the ones found in pure bosonic system [see Table~\ref{z2o3_bosonic} or \ref{o2o2_bosonic} and Table~\ref{o2o3_bosonic}, respectively]. Thus, $O(4)$ and $O(5)$ symmetric MCPs are purely fermion-driven. However, we always find that Yukawa coupling makes the EVs more negative in the subspace of four-boson coupling constants in comparison to those in pure bosonic relativistic theory.
}
\end{figure*}

We also wish to put forward the paramount important role played by massless Dirac fermions in accommodating such high symmetric quantum multi-critical point. We note that a leading order perturbative renormalization group calculation suggests the emergence of $O(2)$ and $O(3)$ symmetric multi-critical points even in pure $Z_2 \otimes Z_2$ and $Z_2 \otimes O(2)$ relativistic bosonic theories respectively. With the addition of gapless fermions, coupled to bosonic order-parameter field via Yukawa coupling, such emergent multi-critical point acquires additional stability [see Figs.~\ref{EV_O2} and ~\ref{EV_O3}]. On the other hand, in a pure bosonic system there exists no stable multi-critical point where $Z_2 \otimes O(3)$ or $O(2) \otimes O(2)$ symmetry gets enlarged to an $O(4)$ symmetry, or $O(2) \otimes O(3)$ symmetry gets enlarged to an $O(5)$ symmetry. Therefore, gapless fermionic excitations are \emph{solely} responsible for giving rise to $O(4)$ and $O(5)$ symmetric quantum multi-critical points and endow them with stability in multi-dimensional coupling constant space [see Figs.~\ref{EV_O4} and ~\ref{EV_O5}]. Some key aspects of emergent quantum multi-critical behavior of a purely bosonic $O(S_1) \otimes O(S_2)$ symmetric field theory are discussed in Appendix~\ref{pure_boson}.

We note that our conclusion regarding the emergence of high symmetric quantum multi-critical points is insensitive to specific choice of the order-parameters, as long as all the competing mass orders mutually anti-commute. In a graphene-like system (accommodating eight component Dirac fermions) there are all together 56 five-tuplets of 5 mutually anti-commuting mass orders~\cite{ruy-hou-mudry-chamon}. All such five-tuplets are unitarily equivalent to the a five-tuplet constituted by either (i) three components of anti-ferromagnet and two components of Kekule valence bond solid or (ii) three components of quantum spin-Hall insulator and two components of $s$-wave pairing~\cite{herbut-lu-roy}. On the other hand, these two five-tuplets of masses can be transformed into each other via a \emph{non-unitary} transformation that, however, leaves the form of the Dirac Hamiltonian unchanged~\cite{herbut-juricic-roy}. Therefore, our conclusions are equally applicable to capture the emergent quantum multi-critical behavior within any five-tuplet (or its subset) of mass orders for two dimensional interacting Dirac fermions.

\begin{table}[t!]
\begin{tabular}{|c|c|c|}
\hline
FPs & $\left( \lambda_1, \lambda_2, \lambda_{12} \right)$ & Eigenvalues of stability matrix \\
\hline \hline
FP$_1$ & $(0,0,0)$ & $(\epsilon, \epsilon, \epsilon)$ \\
\hline
FP$_2$ & $(0,2/3,0)\epsilon$ & $(-\epsilon, \epsilon, 2\epsilon/3)$ \\
\hline
FP$_3$ & $(2/3,0,0)\epsilon$ & $(-\epsilon, \epsilon, 2\epsilon/3)$ \\
\hline
FP$_4$ & $(2/3,2/3,0)\epsilon$ & $(-\epsilon, -\epsilon, \epsilon/3)$ \\
\hline
FP$_5$ & $(3/5,3/5,3/5)\epsilon$ & $(-\epsilon, -4\epsilon/5, -\epsilon/5)$ \\
\hline
FP$_6$ & $(1/3,1/3,1)\epsilon$ & $(-\epsilon, \epsilon/3, 0)$ \\
\hline \hline
\end{tabular}
\caption{ The location of six fixed points (denoted by FP$_j$ for $j=1, \cdots, 6$) in a $Z_2 \otimes Z_2$ symmetric pure bosonic relativistic field theory (with $S_1=S_2=1$). The last column displays the eigenvalues of the stability matrix at various fixed points, suggesting that even in a $Z_2 \otimes Z_2$ symmetric pure bosonic theory, the continuous transition is governed by an $O(2)$ symmetric MCP, namely FP$_5$.
}~\label{z2z2_bosonic}
\end{table}

As a final remark, we note that emergence of high symmetry at a multi-critical point among distinct mass order-parameters that can be rotated into each other by generators of chiral symmetry of Dirac fermions is not specific to two dimensions. For a four-component massless Dirac fermion in three spatial dimensions (a low-energy model for strong spin-orbit coupled Dirac systems, obtained by  neglecting Wilson mass~\cite{roy-goswami-sau}), when the propensities toward the formation of scalar and pseudo-scalar masses become comparable, an $O(2)$ symmetric quantum multi-critical point is realized~\cite{roy-dassarma}. At $d=3$ or $\epsilon=0$ the fixed points are located at $g^2_{1,\ast}=g^2_{2,\ast}=0$ and $\lambda_{1,\ast}=\lambda_{2,\ast}=\lambda_{12,\ast}=0$. Hence, the quantum phase transition from Dirac semimetal to any order phase is of mean-field in nature, captured by the correlation length exponent $\nu=1/2$. The bosonic and fermionic fields become decoupled at the Gaussian critical or multi-critical point, but this process happens very slowly with increasing length scale (logarithmically slow). This logarithmic decoupling causes violation of hyperscaling. However, as shown in Appendix~\ref{Yukawa:3D}, even in $d=3$ if massless Dirac fermions acquire comparable propensities toward formation of two competing ordered phases breaking $O(S_1)$ and $O(S_2)$ symmetries, the ultimate critical behavior is controlled by an $O(S_1+S_2)$ theory. Before flowing to their trivial fixed point values (which is logarithmically slow), the coupling constants from different symmetry sectors approach each other, leading to a larger symmetry at low energies. Furthermore, a complimentary renormalization group calculation about the lower critical one dimension can be performed with $\epsilon=d-1$ as the control parameter~\cite{moshe-moshe}, which immediately shows the existence of an $O(2)$ symmetric multi-critical point~\cite{roy-dassarma}. Following the same route one can demonstrate existence of an $O(6)$ symmetric (maximal symmetry) quantum multi-critical point (in between three-component anti-ferromagnet and three-component valence bond solid in a $\pi$-flux cubic lattice) for three-dimensional, sixteen-component massless Dirac fermions that possess $SU(2)$ spin rotational symmetry~\cite{3D-Dirac:hosur}. We leave that as an exercise for future investigation.

\begin{table}[t!]
\begin{tabular}{|c|c|c|}
\hline
FPs & $\left( \lambda_1, \lambda_2, \lambda_{12} \right)$ & Eigenvalues of stability matrix \\
\hline \hline
FP$_1$ & $(0,0,0)$ & $(\epsilon, \epsilon, \epsilon)$ \\
\hline
FP$_2$ & $(0,2/3,0)\epsilon$ & $(-\epsilon, \epsilon, 2\epsilon/3)$ \\
\hline
FP$_3$ & $(3/5,0,0)\epsilon$ & $(-\epsilon, \epsilon, 3\epsilon/5)$ \\
\hline
FP$_4$ & $(3/5,2/3,0)\epsilon$ & $(-\epsilon, -\epsilon, 4\epsilon/15)$ \\
\hline
FP$_5$ & $(6/11,6/11,6/11)\epsilon$ & $(-\epsilon, -8\epsilon/11, -\epsilon/11)$ \\
\hline
FP$_6$ & $(0.506,0.404,0.69)\epsilon$ & $(-\epsilon, -0.49\epsilon, 0.13 \epsilon)$ \\
\hline \hline
\end{tabular}
\caption{ The location of six fixed points (denoted by FP$_j$ for $j=1, \cdots, 6$) in a $Z_2 \otimes O(2)$ symmetric pure bosonic relativistic theory (with $S_1=1$ and $S_2=2$). The last column displays the eigenvalues of the stability matrix at various fixed points, suggesting that even in a $Z_2 \otimes O(2)$ symmetric pure bosonic theory, the continuous transition is governed by an $O(3)$ symmetric MCP, namely FP$_5$. FP$_6$ can only be found by numerically solving the coupled flow equations.
}~\label{z2o2_bosonic}
\end{table}

\acknowledgments

 B. R. thanks Fakher F. Assaad for useful discussion and Nordita for hospitality. P. G. was supported by the start-up funds from Northwestern University.

\emph{Note added.} During the final stage of preparing this manuscript, we became aware of a similar work where emergent symmetry at a quantum MCP has also been addressed~\cite{lukas-herbut-scherrer}.

\appendix

\section{$O(S_1) \otimes O(S_2)$ Bosonic theory}~\label{pure_boson}

In this appendix we discuss the quantum multi-critical properties of $O(S_1) \otimes O(S_2)$ symmetric, but purely bosonic relativistic field theories. Such analysis can be performed by completely neglecting the boson-fermion coupling in the Euclidean action ${\mathcal S}$, i.e., by setting $g_1=g_2=0$. The RG flow equations for three four-boson couplings, namely $\lambda_1$, $\lambda_2$ and $\lambda_{12}$ can readily be obtained respectively from $\beta_{\lambda_1}$, $\beta_{\lambda_2}$ and $\beta_{\lambda_{12}}$, displayed in Eq.~(\ref{RGFlow_masterequation}), after setting $g^2_1=g^2_2=0$. Our conclusion from the leading of $\epsilon$-expansion are the followings.

1. A $Z_2 \otimes Z_2$ symmetric bosonic $\Phi^4$ theory supports an emergent fully stable (in the critical hyperplane $m^2_1=m^2_2=0$) MCP that possesses an $O(2)$ symmetry. All possible fixed points in such a theory and their stability properties are summarized in Table~\ref{z2z2_bosonic}.

2. A $Z_2 \otimes O(2)$ symmetric bosonic $\Phi^4$ theory accommodates an emergent fully stable (in the critical hyperplane defined by $m^2_1=m^2_2=0$) MCP that possesses an $O(3)$ symmetry. All possible fixed points in such a theory as well as their stability are displayed in Table~\ref{z2o2_bosonic}. However, emergence of such a $O(3)$ symmetric stable fixed point can be an artifact of leading order $\epsilon$-expansion~\cite{calabrese}.

\begin{table}[t!]
\begin{tabular}{|c|c|c|}
\hline
FPs & $\left( \lambda_1, \lambda_2, \lambda_{12} \right)$ & Eigenvalues of stability matrix \\
\hline \hline
FP$_1$ & $(0,0,0)$ & $(\epsilon, \epsilon, \epsilon)$ \\
\hline
FP$_2$ & $(0,6/11,0)\epsilon$ & $(-\epsilon, \epsilon, 6\epsilon/11)$ \\
\hline
FP$_3$ & $(2/3,0,0)\epsilon$ & $(-\epsilon, \epsilon, 2\epsilon/3)$ \\
\hline
FP$_4$ & $(2/3,6/11,0)\epsilon$ & $(-\epsilon, -\epsilon, 7\epsilon/33)$ \\
\hline
FP$_5$ & $(1/2,1/2,1/2)\epsilon$ & $(-\epsilon, -2\epsilon/3, 0)$ \\
\hline \hline
\end{tabular}
\caption{ The location of five fixed points (denoted by FP$_j$ for $j=1, \cdots, 5$) in a $Z_2 \otimes O(3)$ symmetric pure bosonic relativisitc theory (with $S_1=1$ and $S_2=3$). The last column displays the eigenvalues of the stability matrix at various fixed points. Notice that none of the fixed points is fully stable, suggesting a generic first order transition in a $Z_2 \otimes O(3)$ symmetric pure bosonic theory. The fixed point with an emergent $O(4)$ symmetry, namely FP$_5$, has a \emph{marginal} direction, which becomes \emph{relevant} once we incorporate higher order corrections~\cite{calabrese}.
}~\label{z2o3_bosonic}
\end{table}

3. Even though a $Z_2 \otimes O(3)$ symmetric bosonic $\Phi^4$ theory accommodates a fixed point that exhibits an $O(4)$ symmetry in the $m^2_1=m^2_2=0$ hyperplane, this fixed point has a \emph{marginal} direction [see Table~\ref{z2o3_bosonic}]. Such a marginal direction becomes \emph{relevant} once higher order corrections are taken into account~\cite{calabrese}. A qualitatively similar situation also arises in an relativistic $O(2) \otimes O(2)$ symmetric bosonic $\Phi^4$ theory [see Table~\ref{o2o2_bosonic}].

\begin{table}[t!]
\begin{tabular}{|c|c|c|}
\hline
FPs & $\left( \lambda_{1}, \lambda_{2}, \lambda_{12} \right)$ & Eigenvalues of stability matrix \\
\hline \hline
FP$_1$ & $(0,0,0)$ & $(\epsilon, \epsilon, \epsilon)$ \\
\hline
FP$_2$ & $(0,3/5,0)\epsilon$ & $(-\epsilon, \epsilon, 3\epsilon/5)$ \\
\hline
FP$_3$ & $(3/5,0,0)\epsilon$ & $(-\epsilon, \epsilon, 3\epsilon/5)$ \\
\hline
FP$_4$ & $(3/5,3/5,0)\epsilon$ & $(-\epsilon, -\epsilon, \epsilon/5)$ \\
\hline
FP$_5$ & $(1/2,1/2,1/2)\epsilon$ & $(-\epsilon, -2\epsilon/3, 0)$ \\
\hline \hline
\end{tabular}
\caption{ The location of five fixed points (denoted by FP$_j$ for $j=1, \cdots, 5$) in a $O(2) \otimes O(2)$ symmetric pure bosonic theory (with $S_1=S_2=2$). The last column displays the eigenvalues of the stability matrix at various fixed points.  There is no fully stable fixed point in the three-dimensional coupling constant space $\left(\lambda_1, \lambda_2, \lambda_{12} \right)$ in a pure bosonic theory. Notice that the fixed point FP$_5$ features one marginal direction, similar to the situation for $Z_2 \otimes O(3)$ theory [see Table~\ref{z2o3_bosonic}].
}~\label{o2o2_bosonic}
\end{table}

4. Finally, from the leading order $\epsilon$-expansion of a $O(2) \otimes O(3)$ symmetric bosonic $\Phi^4$ theory we find that there exists an $O(5)$ symmetric fixed point in the $m^2_1=m^2_2=0$ hyperplane. However, such high symmetric fixed point is not fully stable and has one \emph{relevant} direction. Stability of all the fixed points in such a bosonic theory is shown in Table~\ref{o2o3_bosonic}.

\begin{table}[t!]
\begin{tabular}{|c|c|c|}
\hline
FPs & $\left( \lambda_1, \lambda_2, \lambda_{12} \right)$ & Eigenvalues of stability matrix \\
\hline \hline
FP$_1$ & $(0,0,0)$ & $(\epsilon, \epsilon, \epsilon)$ \\
\hline
FP$_2$ & $(0,6/11,0)\epsilon$ & $(-\epsilon, \epsilon, 6\epsilon/11)$ \\
\hline
FP$_3$ & $(3/5,0,0)\epsilon$ & $(-\epsilon, \epsilon, 3\epsilon/5)$ \\
\hline
FP$_4$ & $(3/5,6/11,0)\epsilon$ & $(-\epsilon, -\epsilon, 0.14\epsilon)$ \\
\hline
FP$_5$ & $(6/13,6/13,6/13)\epsilon$ & $(-\epsilon, -0.615\epsilon, 0.077\epsilon)$ \\
\hline
FP$_6$ & $(0.542,0.5085,0.32)\epsilon$ & $(-\epsilon, -0.83 \epsilon, -0.05 \epsilon)$ \\
\hline \hline
\end{tabular}
\caption{ The location of six fixed points (denoted by FP$_j$ for $j=1, \cdots, 6$) in a $O(2) \otimes O(3)$ symmetric pure bosonic theory (with $S_1=2$ and $S_2=3$). The last column displays the eigenvalues of the stability matrix at various fixed points. FP$_{6}$ can only be found by numerically solving the coupled flow equations. Notice that $O(5)$ symmetric fixed point (namely FP$_5$) is not a critical one in a pure bosonic theory, a mixed fixed point, namely FP$_6$, stands as a critical point.
However, we note that the existence of the mixed QCP (FP$_6$), obtained here from the leading order $\epsilon$-expansion, could be deceptive, as the ultimate stable fixed point is believed to be one of the decoupled ones (namely FP$_2$ or FP$_3$)~\cite{calabrese}.
}~\label{o2o3_bosonic}
\end{table}

To summarize, we find that only $Z_2 \otimes Z_2$ and $Z_2 \otimes O(2)$ symmetric purely bosonic relativistic $\Phi^4$ theories respectively support fully stable $O(2)$ and $O(3)$ symmetric quantum MCP (at least from the leading order $\epsilon$-expansion). Inclusion of Yukawa couplings further enhances the stability of these two MCPs. Such an outcome can be seen from the scaling of the \emph{negative} eigenvalues of the stability matrix in the $\left( \lambda_1, \lambda_2, \lambda_{12} \right)$ ~plane, as shown in Figs.~\ref{EV_O2} and ~\ref{EV_O3}. By contrast, a $Z_2 \otimes O(3)$ [or $O(2) \otimes O(2)$] and $O(2) \otimes O(3)$ symmetric relativistic bosonic theories do not support stable $O(4)$ or $O(5)$ symmetric MCP. Therefore, emergence of $O(S)$ symmetric fully stable MCP with $S\geq 4$ in a GNY model can solely be attributed to the non-trivial coupling between bosonic and fermionic degrees of freedom. This outcome can also be visualized from the scaling of the eigenvalues of the stability matrix in $\left( \lambda_1, \lambda_2, \lambda_{12} \right)$ ~plane, as shown in Figs.~\ref{EV_O4} and ~\ref{EV_O5}.


\section{Restoration of largest symmetry at multi-critical point}~\label{Sec:largestsymm}

A GNY field theory describing a competition among several orders may feature MCPs with different symmetries, and the question is which one of the possible MCPs ultimately governs the quantum multi-critical behavior. We address this problem in a specific example with many such possibilities, and which, together with the ones studied in previous sections, leads us to establish that \emph{if interacting Dirac fermions are conducive to multiple orderings, the resulting MCP restores largest possible symmetry.} Although the following field theory has previously been considered in Ref.~\cite{roy-juricic-mcp}, in the context of present discussion it is worth revisiting this problem.

An example of such scenario is provided by an effective field theory with the Lagrangian $L=L_f+L_{Y}+ L_{b}$, where $L_f=\bar{\Psi}(x) \gamma_\mu \partial_\mu \Psi(x)$, $L_{Y}=L^s_{Y}+L^t_{Y}$ with
\begin{eqnarray}
L^{s}_{Y} &=& g_{1s} \varphi \; \bar{\Psi}(x) \Psi(x) + g_{2s} \chi \; \bar{\Psi}(x) i \gamma_5 \Psi(x), \\
L^{t}_{Y} &=& g_{1t} \vec{\varphi} \cdot \bar{\Psi}(x) \vec{\sigma} \Psi(x) + g_{2t} \vec{\chi} \cdot \bar{\Psi}(x) i \vec{\sigma} \gamma_5 \Psi(x).
\end{eqnarray}
The bosonic part of the Lagrangian can be decomposed according to $L_b=L^s_b+L^t_b+L^{s-t}_b$, with
\begin{eqnarray}
L^s_b = \sum_{\alpha=\varphi, \chi} \left[ \frac{1}{2} \left( \partial_\mu \alpha \right)^2 + m^2_\alpha \alpha^2 + \frac{\lambda_\alpha}{4!} \alpha^4 \right] + \frac{\lambda_{\varphi \chi}}{12} \varphi^2 \chi^2, \nonumber \\
\end{eqnarray}
\begin{equation}
L^t_b = \sum_{\alpha=\vec{\varphi}, \vec{\chi}} \left[ \frac{1}{2} \left( \partial_\mu \vec{\alpha} \right)^2 + m^2_{\vec{\alpha}} \vec{\alpha}^2 +   \frac{\lambda_{\vec{\alpha}}}{4!} \left( \vec{\alpha} \cdot \vec{\alpha} \right)^2 \right] + \frac{\lambda_{\vec{\varphi} \vec{\chi}}}{12} \vec{\varphi}^2 \vec{\chi}^2,
\end{equation}
\begin{equation}
L^{s-t}_b=\sum_{\alpha=\varphi, \chi} \frac{\lambda_{\alpha \vec{\alpha}}}{12} \alpha^2 \vec{\alpha}^2 + \frac{\lambda_{\varphi \vec{\chi}}}{12} \varphi^2 \vec{\chi}^2 + \frac{\lambda_{\vec{\varphi} \chi}}{12} \chi^2 \vec{\varphi}^2.
\end{equation}
The above coupled field theory with appropriate choice of spinor basis $\Psi$ and the representation of mutually anti-commuting four-component $\gamma$-matrices can describe the following scenarios: (a) competition between spin-singlet (described by $\varphi, \chi$) and spin-triplet (described by $\vec{\varphi}, \vec{\chi}$) KVBS or (b) competition between spin-singlet $s$-wave pairing (with $\varphi$ and $\chi$ respectively representing its real and imaginary components) and spin-triplet $f$-wave (with $\vec{\varphi}$ and $\vec{\chi}$ respectively representing its real and imaginary components) pairing. However, the outcome of this analysis is impervious to these microscopic details.

Before we proceed with the RG analysis of the above model we annouce possible outcomes regarding MCPs:

1. A $Z_2$ symmetric MCP with only $g^2_{1s} \sim \epsilon$, but $g^2_{2s}=g^2_{1t}=g^2_{2t}=0$, which describes a transition to an ordered phase $\langle \varphi \rangle \neq 0$, but $\langle \chi \rangle= \langle \vec{\varphi} \rangle =\langle \vec{\chi} \rangle= 0$.

2. An $O(2)$ symmetric MCP with, for example, $g^2_{1s}=g^2_{2s} \sim \epsilon$, but $g^2_{1t}=g^2_{2t}=0$ that described a continuous transition to an ordered phase where $\langle \varphi \rangle, \langle \chi \rangle \neq 0$, but $\langle \vec{\varphi} \rangle =\langle \vec{\chi} \rangle= 0$.

3. An $O(3)$ symmetric MCP, with $g^2_{1t} \sim \epsilon$, but $g^2_{1s}=g^2_{2s}=g^2_{2t}=0$. Such MCP describes a continuous transition to an ordered phase where $\langle \vec{\varphi} \rangle \neq 0$, but $\langle \varphi \rangle=\langle \chi \rangle=\langle \vec{\chi} \rangle= 0$.

4. An $O(4)$ symmetric MCP, with $g^2_{1s}=g^2_{2t} \sim \epsilon$ and $g^2_{2s}=g^2_{1t}=0$ or vice-versa. Such MCP can be responsible for nucleation of an ordered phase with $\langle \varphi \rangle, \langle \vec{\chi} \rangle \neq 0$, but $\langle \chi \rangle, \langle \vec{\varphi} \rangle \neq 0$.

Therefore, the above theory allows us to investigate an interesting issue; emergence of ultimate MCP in the presence of multiple possibilities with a different symmetries.

\begin{figure}[t!]
\subfigure[]{
\includegraphics[width=4cm,height=3.75cm]{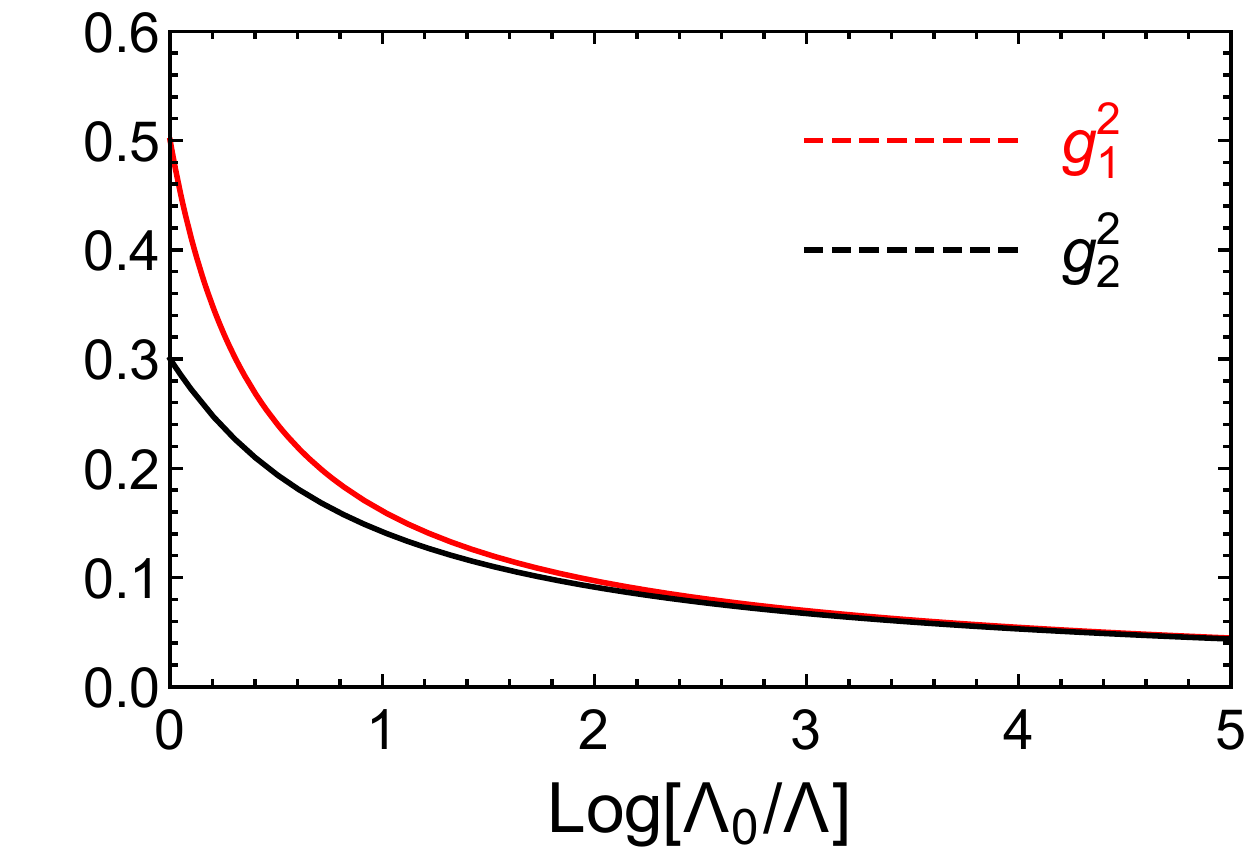}
}
\subfigure[]{
\includegraphics[width=4cm,height=3.75cm]{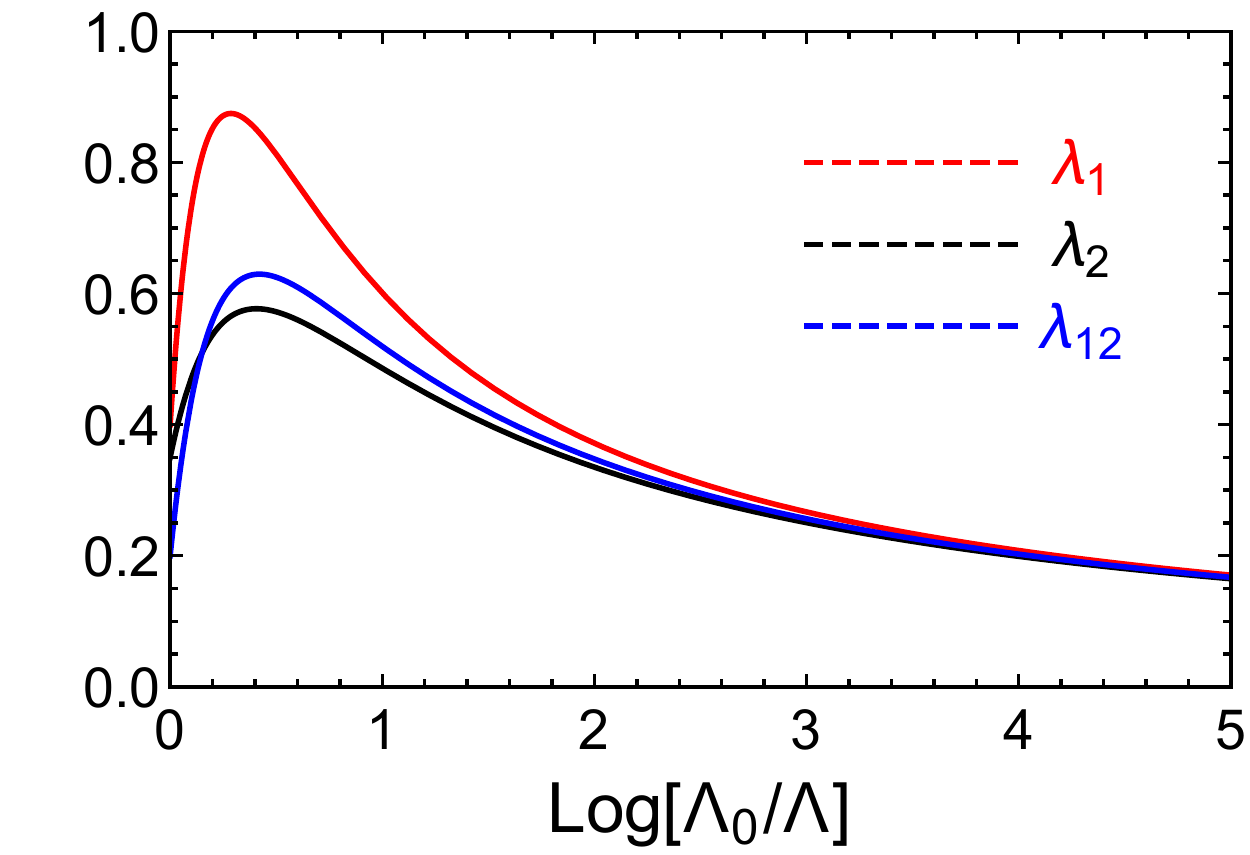}
}
\caption{  Renormalization group flow of (a) Yuakwa couplings and (b) four-boson couplings in three spatial dimensions of an $O(1) \otimes O(1)$ symmetric GNY model, with $N=1$. Even though all the couplings are flowing to zero logarithmically slow, prior to reaching trivial fixed point values they all acquire identical values, indicating emergence of an $O(2)$ symmetry at a Gaussian MCP. Here $\log[\Lambda_0/\Lambda]$ represents running RG time.
}~\label{flow_3D}
\end{figure}

Now we proceed with the RG analysis to the leading order in the $\epsilon$-expansion. Since we have seen so far that at least to the leading order the sub-space of Yukawa couplings is decoupled from four-boson couplings, we first focus on the four-dimensional coupling constant space spanned by $g_{1s}$, $g_{2s}$, $g_{1t}$ and $g_{2t}$. To the leading order in $\epsilon$-expansion the coupled RG flow equations read as
\allowdisplaybreaks[4]
\begin{eqnarray}
\beta_{g^2_{1s}} &=& \epsilon g^2_{1s} -(2N+3) g^4_{1s} + g^2_{1s} \left( g^2_{2s} - 9 g^2_{1t} + 3 g^2_{2t} \right), \nonumber \\
\beta_{g^2_{2s}} &=& \epsilon g^2_{2s} -(2N+3) g^4_{2s} + g^2_{2s} \left( g^2_{1s} - 9 g^2_{2t} + 3 g^2_{1t}  \right), \nonumber \\
\beta_{g^2_{1t}} &=& \epsilon g^2_{1t} -(2N+1) g^4_{1t} - g^2_{1t} \left(5 g^2_{2t} + 3 g^2_{1s} -g^2_{2s} \right),  \nonumber \\
\beta_{g^2_{2t}} &=& \epsilon g^2_{2t} -(2N+1) g^4_{2t} - g^2_{2t} \left(5 g^2_{1t} + 3 g^2_{2s} -g^2_{1s} \right). \nonumber \\
\end{eqnarray}
These coupled RG flow equations support only two fully stable fixed points, located at
\allowdisplaybreaks[4]
\begin{eqnarray}
A: \; g^2_{1s,\ast}=g^2_{2t,\ast}=\frac{\epsilon}{2N}, g^2_{2s,\ast}=g^2_{1t,\ast}=0, \nonumber \\
B: \; g^2_{2s,\ast}=g^2_{1t,\ast}=\frac{\epsilon}{2N}, g^2_{1s,\ast}=g^2_{2t,\ast}=0. \nonumber
\end{eqnarray}
Notice that upon substituting the fixed point values of $A$ or $B$, the Yukawa part of the Lagrangian $L_{Y}$ assumes an $O(4)$ symmetric form. Therefore, out of all possible QCPs, the system selects the ones which possess the largest symmetry, $O(4)$ in our case. If we now consider only one of the two fixed points, say $A$, the bosonic part of the Lagrangian simplifies enormously as we enjoy the liberty of setting
\begin{eqnarray}
\lambda_\chi = \lambda_{\varphi \chi}= \lambda_{\vec{\varphi}}= \lambda_{\vec{\varphi} \vec{\chi}}=\lambda_{\chi \vec{\varphi}}=\lambda_{\chi \vec{\chi}}=0
\end{eqnarray}
from outset. The RG flow equations for $\lambda_\varphi$, $\lambda_{\vec{\chi}}$ and $\lambda_{\varphi\vec{\chi}}$ can readily be obtained from Eq.~(\ref{RGFlow_masterequation}), taking $g^2_{1} \to g^2_{1s}$, $g^2_{2} \to g^2_{1t}$, $\lambda_1 \to \lambda_\varphi$, $\lambda_2 \to \lambda_{\vec{\chi}}$ and $\lambda_{12} \to \lambda_{\varphi \vec{\chi}}$, with $S_1=1$ and $S_2=3$. Rest of the analysis follows the identical steps outlined in Sec.~\ref{Sec:generalMCP}, which ensure the existence of an $O(4)$ symmetric quantum MCP.

Note that among all four possible MCPs, possessing $Z_2$ (with $S=1$), $O(2)$ (with $S=2$), $O(3)$ (with $S=3$) and $O(4)$ (with $S=4$) symmetries, the correlation length exponent ($\nu^{(S)}$) [see Eq.~(\ref{CLE:final})], fermionic anomalous dimension ($\eta^{(S)}_\Psi$) [see Eq.~(\ref{anomalousdim:Fer})] and bosonic anomalous dimension ($\eta^{(S)}_b$) [see Eq.~(\ref{anomalousdim:Bos})] is largest at the $O(4)$ symmetric MCP. Therefore, the above exercise allows us to further strengthen the proposed conjecture regarding the nature of the ultimate quantum MCP when multiple phases compete with each other: \emph{The ultimate quantum MCP possesses the largest bosonic and fermionic anomalous dimensions, and also the largest correlation length exponent}.

\section{Emergent symmetry in $(3+1)$ dimensions}~\label{Yukawa:3D}

In this appendix we provide some quantitative aspect of emergent symmetry among competing orders for strongly interacting three-dimensional Dirac fermions. The effective field theory assumes the same form as the one shown in Sec.~\ref{Sec:generalMCP}, and the RG flow equations can readily be obtained from Eq.~(\ref{RGFlow_masterequation}) after setting $\epsilon=0$, since now we are at $d=3$. Various scenarios can be studied by specific choices of the parameters $N$, $S_1$ and $S_2$. For example, (a) setting $S_1=1=S_2$ and $N=1$ (single four component Dirac fermions in strong spin-orbit coupled model) we can address the emergent quantum multi-critical behavior when propensity toward scalar and pseduo-scalar mass orderings are comparable, or (b) the compeition between three-dimensional anti-ferromagnet and valence bond solid in a Diac system possessing $SU(2)$ spin rotational symmetry can be addressed by setting $S_1=S_2=3$ and $N=4$. Irrespective of these details the emergent quantum critical phenomena in different realization of Dirac fermions share some common features in $d=3$:

1. All the transitions take place for $g^2_{\ast,j}=0$ and $\lambda_{\ast,j}=0$, and consequently the correlation length exponent $\nu$ is locked to its mean-field value $\nu=1/2$. As a result physical observables display \emph{logarithmic violation of scaling}.

2. The fermionic and bosonic anomalous dimensions are exactly zero (i.e., $\eta_f=\eta_b=0$), also indicating that these two degrees of freedom are decoupled from each other near a quantum phase transition~\cite{generalWeyl}.

However, the flows of $g^2_j$s and $\lambda_j$s toward their trivial fixed point values are \emph{logarithmically slow}. When massless Dirac fermions acquire comparable propensity toward the formation of distinct mass orders that break $O(S_1)$ and $O(S_2)$ symmetries in the ordered phases, but can be rotated into each other by generators (exact or emergent) of chiral symmetry of Dirac fermions, then one can show that all Yukawa and four-boson couplings flow toward same value (thus emergence of $O(S_1+S_2)$ or $O(S)$ symmetry) prior to end up at trivial fixed point value. This scenario has been demonstrated in Fig.~\ref{flow_3D}.

\end{document}